\documentclass[preprint,amsmath,amssymb,prb,aps]{revtex4}
\usepackage{graphicx}
\usepackage{amsmath,amssymb}
\usepackage{mathrsfs}
\usepackage{color}
\usepackage{afterpage}
\usepackage{enumerate}
\usepackage[version=3]{mhchem}
\usepackage[normal]{subfigure}

\begin{document}
\title{Efficient {\em ab initio} schemes for finding thermodynamically stable and metastable atomic structures: Benchmark of cascade genetic algorithms.}
\author{Saswata Bhattacharya, Sergey V. Levchenko, Luca M. Ghiringhelli, and Matthias Scheffler}
\affiliation{Fritz-Haber-Institut der Max-Planck-Gesellschaft, Faradayweg 4-6, D-14195 Berlin, Germany}
\date{\today}
\begin{abstract}
A first-principles based methodology for efficiently and accurately finding thermodynamically stable and metastable atomic structures
is introduced and benchmarked.
The approach is demonstrated for gas-phase metal-oxide clusters in thermodynamic equilibrium with a reactive (oxygen) atmosphere at finite pressure and temperature. It consists of two steps.
At first, the potential-energy surface is scanned by means of a global-optimization technique, i.e., a massive-parallel first-principles cascade genetic algorithm for which the choice of all parameters is validated against higher-level methods. In particular, we validate a) the criteria for selection and combination of structures used for the assemblage of new candidate structures, and b) the choice of the exchange-correlation functional. The selection criteria are validated against a fully unbiased method: replica-exchange molecular dynamics. Our choice of the exchange-correlation functional, the van-der-Waals-corrected PBE0 hybrid functional, is justified by comparisons up to highest level currently achievable within density-functional theory, i.e., the renormalized second-order perturbation theory, rPT2.
In the second step, the low-energy structures are analyzed by means of {\em ab initio} atomistic thermodynamics in order to determine compositions and structures that minimize the Gibbs free energy at given temperature and pressure of the reactive atmosphere.
\end{abstract}
\maketitle
\section{Introduction}
\label{intro}
A functional material, such as a solid-state catalyst, is in general very different from the pristine material which is initially introduced into the reactive environment.
After a so called ``induction period'', nanostructures of various shapes and compositions, point defects, extended defects such as steps, dislocations, and stacking faults, can result from and will be modified by interaction of the surface of the catalyst with the reactive environment. Therefore, for a functional material design, it is crucial to reliably predict stoichiometry and structure of the material under realistic conditions. However, the compositional and structural degrees of freedom at the material-gas interface result in an enormous (combinatorial) increase of the configurational space.

Materials at nanoscale size show unusual properties, that have prompted extensive studies of clusters \cite{new1, r5, new2, new3, new4, r1} and that are generally attributed to the presence of undercoordinated atoms.
However, in the presence of a reactive atmosphere, clusters can adsorb species from the gas phase, changing their stoichiometry. Under certain conditions this new state, not the initial cluster, can become the active (functional) material. Thus, in order to understand the functional properties of clusters in a reactive atmosphere, it is important to know which structures and stoichiometries are energetically accessible.

In this paper, we introduce a robust first-principles methodology for the determination of (meta)stable structures at realistic environmental conditions. In particular, we address the problem of validation of methods for global structure optimization. No existing method provides criteria for identifying a local minimum on a potential-energy surface as a global one. We show how this problem can be addressed, and describe a methodology for selecting a structure optimization scheme that is more likely
to find the global energy minimum for a given system with less iterations.

The paper proceeds as follows. The first step is an extensive and efficient scanning of the potential-energy surface (PES) by global structure optimization; subsequently, the influence of the experimental conditions (here, temperature and pressure of a reactive atmosphere) is included. As model system, we study gas-phase Mg$_M$O$_x$ clusters, where $M = 1,2,3,10, 18$ (see Ref. \cite{letter} for results on this system at other values of $M$) and $x$ is the composition which minimizes the cluster free energy at given environmental conditions.
Up to now, different methodologies for global structure optimization and levels of theory for the inter-atomic interactions have been employed to study only stoichiometric, (MgO)$_M$, clusters.
\cite{r5,letter,r1,r2,r3,r4,r6,r7,r8,r9,r10,r11,r12,r13,r14a,r14b,r14c,r14d,r14e,r14f,r14g,r14h,r14i,r14j,r14k,r14l,r14m,r14n,r14o}
The previous studies employed \textit{ab initio} or empirical potentials for the evaluation of the total and relative energies. Both approaches have certain advantages and drawbacks. Calculations based on empirical potential or force fields are computationally cheap and fast enough to thoroughly scan the PES, but results greatly depend upon the choice of the parameters used\cite{r7} (see our analysis and discussion below). In contrast, \textit{ab initio} calculations entail a fully quantum-mechanical description of the chemical bonding and of bond breaking/forming. Thus, \textit{ab initio} calculations are generally more accurate. However they are computationally expensive and the accuracy of the results depends on the exchange-correlation functional\cite{r14e}, which must be approximated. This approximation must be validated against higher-level methods.

As it will become clear along the paper, the model system that we chose is representative of a class of systems for which very accurate evaluation of the potential-energy surface (PES) is mandatory for meaningful predictions (for instance, we will demonstrate that hybrid exchange-correlation functionals are required). Since high-accuracy calculations are computationally very expensive, a scheme for an efficient scanning of the PES is necessary in order to avoid wasting time with high-accuracy calculations in uninteresting regions of the configurational phase space.

The first step of our methodology, i.e., the PES scanning, was chosen to be performed via a massive-parallel cascade genetic algorithm (cGA).
In such global scan, we look for the set of coordinates (geometrical structure) {\em as well as spins} that minimize the total energy of the cluster.
An alternative scheme, where parallel searches at fixed spins are performed is discussed in section \ref{sec:spin}.
The term ``cascade" is used because the global-optimization is designed as a multi-step procedure involving increasing levels of accuracy for the evaluation of the globally-optimized quantity, which, in our case, is total energy.

In section \ref{pgs} we introduce our cGA scheme. The method is not unique, as many choices are possible at every sub-step of its implementation. Therefore, we propose a validation scheme for the identification of those crucial parameters that need a careful tuning in order to achieve an efficient and accurate estimate of the desired quantities (e.g., structure and energy of the global minimum at each cluster size). The details of this validation, which constitute the core message of this paper, are described in sections \ref{ga-remd} and \ref{comp-funct}.

As second step, the structures found via the cGA are post-processed by applying the concepts of \textit{ab initio} atomistic thermodynamics (\textit{ai}AT) \cite{aiat} in order
to analyze the $(T,p_{\textrm{O}_2})$ dependence of the stable compositions and structures of the various sizes $M$ of the Mg$_M$O$_x$ clusters in the oxygen atmosphere. A first-principles description of the environmental effects on materials has been previously developed and appplied extendively for predicting the relative stabilities of different phases in bulk semiconductors and surface of semiconfductors and oxides.\cite{ms1,ms2,ms3,ms4,ms5} In section \ref{sec:aiat} we present a self-contained introduction of \textit{ai}AT, here adapted to the study of the relative stabilities of small Mg$_M$O$_x$ clusters at a finite temperature and pressure.\cite{lmg1}.

After showing the phase diagram of Mg$_2$O$_x$, we discuss the issue of ``coexistence regions''. By coexistence regions, we mean areas of the $(T,p)$ phase diagram where different compositions and/or different isomers of the same composition coexist, as their free energies are nearly degenerate.

\section{Theory and Methods}
\label{methods}
\subsection{Parallel Global Search}
\label{pgs}

We have implemented a parallel algorithm for global search, tailored for cluster structures. Building on the known genetic algorithm (GA)\cite{r15,r15a,r16,r17} approach, we designed a cascade genetic algorithm (cGA) for finding the global-minimum- (GM) and lowest local-minimum-energy structures of a cluster of certain size and composition.

GA is a global-optimization technique based on the principles of natural evolution. In general, a GA includes the following steps: An initial population is formed with a group of individuals created randomly. The individuals in the population are then evaluated via a so-called fitness function, designed to quantify how well they satisfy chosen criteria. This (scalar) fitness is the quantity to be optimized. Two individuals are then randomly selected, with weight based on their fitness, the higher the fitness, the higher the chance of being selected. These individuals ``mate'', i.e., are combined to create one candidate offspring, that can be in turn ``mutated'' randomly. In the next selection for ``mating'', this new individual is included in the pool of candidates and can be selected on the basis of its fitness . This continues until a suitable solution has been found or a certain number of iterations have passed, depending on convergence criteria.

In our cascade scheme successive steps employ higher level of theory and each of the next level takes information obtained at the immediate lower level. It proceeds by first performing an extensive GA pre-scanning of structures
 by means of a computationally inexpensive classical force field (here, ReaxFF \cite{ff,ff1}). The low-energy structures found with this scanning are used as initial guess for further, extensive GA for the minimization of the \textit{ab initio} total energy of the clusters. In this second GA search, we look for global minima (at different stoichiometries) of the PES's described by the the van-der-Waals-corrected\cite{ts-scheme} PBE0\cite{pbe0} hybrid functional (PBE0+vdW). In order to alleviate the computationally (prohibitively) expensive direct relaxation (local optimization) of structures with a hybrid functional, we found that a very efficient strategy is to relax the candidate structures by means of a lower-level functional, namely PBE+vdW (the van-der-Waals-corrected \cite{ts-scheme} PBE \cite{pbe} functional) but the PBE+vdW energetics are not used for the evaluation of {\em fitness} function. The {\em fitness} function is evaluated after finishing an additional step of calculating total energy using PBE0+vdW at the geometry fixed to the one found in the former step (see below for details and discussion of an alternative scheme). Such an algorithm yields reliable predictions only if local equilibrium PBE+vdW structures are close to hybrid xc-functional one. This assumption has been thoroughly tested for selected Mg$_M$O$_x$ clusters (namely, Mg$_{10}$O$_{10}$,  MgO$_{4}$,  Mg$_{2}$O$_{2}$,  Mg$_{2}$O$_{5}$,  Mg$_{5}$O$_{8}$)
where we found that the difference between PBE0+vdW binding energies for structures relaxed at the GGA xc-functional level and that of at the hybrid xc-functional level are negligible.


One major possible source of inefficiency in using a DFT-based search from scratch is related to the creation of the initial pool. Since the aim of the GA approach is to perform an unbiased scan of the structures, the initial pool should be ideally composed of (constrained \footnote{In the practice, the coordinates of the atoms composing the structure are not fully randomly selected, but some constrains have to be posed, the obvious one being that the atoms are ``not too close'', {\em vide infra} for details according to our implementation.}) random structures at a given composition.
However, running a geometrical optimization by starting form a random structure can require several hundreds if not thousands of optimization steps. Performing such optimization fully at the DFT level is impractical. Furthermore, a lot of computational effort may then be spent in converging the forces for configurations very far from a local minimum, where high accuracy is not needed.
A way of overcoming this problem is to perform local optimization of randomly formed structures by means of a force field. In principle this pre-relaxation could serve already as a starting point for DFT-based GA. In practice, due to the inexpensiveness of a classical force field when compared to DFT calculations, we performed a thorough GA scan with the force field and the low-energy structures generated in this way served as {\em initial pool} for the DFT scan. This choice, however, carries a potential danger, i.e., that the optimization performed at the force-field (FF) level biases the initial pool for the DFT scan, such that a relevant portion of the configurational phase space remains unexplored. In the following we show that, when using a FF as flexible as reaxFF for the FF-based initial GA, no such bias is introduced. However, we also show that the reaxFF results are subject to significant errors in the relative energies and structures. Even if a DFT post-relaxation is performed on the pool obtained via reaxFF, important structures (including the GM) can be missed.

Schematically, our cGA algorithm proceeds as follows (all terms typeset in italic will be explained afterwards):
\begin{enumerate}  [(1)]
 \item Selection of a composition of the clusters and formation of an initial pool of (constrained) random structures, locally optimized by a FF.
 \item Evaluation of the {\em fitness} function for all structures (using FF binding energy). 
 \item GA global optimization using the FF. This consists of the following steps (i)--(v), which are iterated until {\em convergence}.
 \begin{enumerate}[(i)]
  \item {\em Selection} of two structures (in GA jargon, {\em parents}).
  \item Assemblage of a new trial structure ({\em child}) through {\em crossover} and {\em mutation}.
  \item Local optimization (force minimization) of the child structure using the FF.
  \item Evaluation of the {\em fitness} function. Comparison of the optimized child with existent structures, reject if {\em similar}: Jump to (i).
  If not rejected, the next step is performed.
  \item A check whether {\em convergence} has been reached. If so, FF-GA is stopped and the next step, i.e., DFT-GA, is performed.
 \end{enumerate}
 \item Formation of a new pool of structures using structures with highest fitness from FF-GA, locally optimized at the DFT level (PBE+vdW, {\em low-level settings}).
 \item Calculation of fitness function for all structures (using energy at the PBE0+vdW level).
 \item GA scheme using DFT. In practice, iteration of steps (a)--(i):
 \begin{enumerate}[(a)]
  \item {\em Selection} of two structures.
  \item Assemblage of a child structure through {\em crossover} and {\em mutation}.
  \item Local optimization of the child structure with PBE+vdW, {\em low-level settings}.
  \item Comparison of the optimized child with existent structures. {\em Early rejection} if {\em similar}: Jump to (a).
  If not rejected, the next step is performed.
  \item Further local optimization of the child with PBE+vdW, {\em high-level settings}.
  \item Evaluation of {\em fitness} function based on PBE0+vdW total energy, with the geometry found in the former step.
  \item Check whether {\em convergence} has been reached. If so, stop.
 \end{enumerate}
\end{enumerate}

In the following, we explain one by one the keywords introduced in the scheme above.

\subsubsection{\bf Initial random pool}
\label{rp}
 We generate structures with atoms randomly distributed on the surface of an ellipsoid with axes of random length, with the constraint that the closest distance between neighbours is larger than a certain threshold value and for Mg-O, O-O, and Mg-Mg, the threshold values were set to 1.5, 1.0 and 2.75 $\textrm{\AA}$, respectively. These threshold values are set as 75\% of the equilibrium bond distance of the three isolated dimers. The constrained random generation algoritm is built in order to get structures in three flavours: nearly spherical (three axes of almost equal length), prolate (two short axes, one long) and oblate (two long axes and one short). The limit of prolate structures are linear structures and the limit of oblate ones are planar structures.\\ For benchmarking, we strictly obeyed these rules. However, in practice it is sometimes beneficial to accompany the random structures with some other structures constructed by using any available prior knowledge or chemical intuition of the system. One (obvious) example in the case of the Mg$_M$O$_x$ clusters is constructing a parallelepipedal rocksalt-like structure for a stoichiometric cluster with a suitable number of atoms (e.g., $2\times2\times2$, $3\times4\times5$ ...).

\subsubsection{\bf Fitness function}
\label{ff}
Each cluster $i$ in the population is assigned a normalized fitness value, $\rho_i$, based on its total energy (binding energy for the FF):
\begin{equation}
\label{eqn1}
\rho_i=\frac{\epsilon_i}{\sum_i\epsilon_i}
\end{equation}
and $\epsilon_i$ is the relative energy of the $i^{th}$ cluster as defined below:
\begin{equation}
\label{eqn2}
\epsilon_i=\frac{E_\textrm{max}-E_i}{E_\textrm{max}-E_\textrm{min}}
\end{equation}
Where $E_i$ is the total energy of the $i^{th}$ cluster of the population and $E_\textrm{min}, E_\textrm{max}$ correspond to the dynamically updated lowest and highest total energies in the population, respectively.

With this definition, low (more negative) energy clusters have high fitness and high (less negative) energy clusters have low fitness.

\subsubsection{\bf Selection rule}
\label{sr}
 We use a ``roulette-wheel'' selection criterion \cite{roulette} with selection probability proportional to the value of the normalized fitness function. The idea is that the lower the total (or binding) energy (i.e., large negative value) of a certain configuration, the larger the probability to be chosen from the population. A cluster is picked at random and is selected for mating if its normalized fitness value ($\rho_i$, defined in the previous paragraph) is greater than $\textrm{Rand}[0,1]$, a randomly generated number, uniform in the interval $[0,1]$.

A subtle problem is related to a possible ``poisoning'' of the selection pool with many structures that are all similar too each other. We have noticed that, frequently, a basin in the PES contains many local minima. These minima are different enough from each other to be judged as not {\em similar} by the geometric criterion defined below; on the other hand, some persistent topological feature is shared among all such minima. In such cases, the genetic pool may be flooded by a large number of alike structures, energetically close to the running GM, due to the high likelihood that mating among similar structures produces similar structures.
GA takes then significantly long time to reach another basin in the PES. This problem is known as ``too early convergence'' (or pre-convergence) \cite{r16} and, traditionally, the only suggested strategy to obviate this problem is to restart GA using a set of completely different random initial structures. However, we found (see below) that by selecting with small but non-negligible probability one ``bad'' (high-energy) structure in the population helps in moving out to a different basin. Therefore, we define also a complementary fitness function $\tilde{\rho}_j = (1-\rho_j)$ so that, with properly tuned frequency (see section \ref{ga-remd} for details), we allow for the selection of one structure with high $\tilde{\rho}_j$, which is combined via {\em crossover} with another structure selected via the usual high-$\rho_i$ criterion. In section \ref{ga-remd}, we show that this choice greatly helps the convergence of the GA scheme and we also show how we optimized the mixing ratio among different selection rules.

\subsubsection{\bf Crossover}
\label{sec:cross}


The crossover operator takes care of combining the parent clusters selected as explained above.
We have implemented three rather different crossover operators. All such operators are applied to two (one could use more than two, in principle) selected parents and the first step is always to apply to both parent clusters a random rotation around their center of geometry, which is fixed at the origin of the coordinate axes for convenience.
The crossover schemes differ in the way the hard constraint of imposing to the newly assembled cluster (the {\em child}) the same stoichiometry of the parent clusters is implemented.

(i) Crossover-1: This is a combined crossover and mutation (see below).
The strategy is to decouple the atomic coordinates from their species. Let us consider that atomic coordinates of a cluster are listed in a ``geometry'' file where each line contains the four coordinates (three space plus one spin coordinate) of an atom and a label for its species. The sequence of atomic species in such geometry file is defined once and for all throughout the GA scan. This fixes the stoichiometry. If the cluster contains an even number $N$ of atoms ($N=M+x$), a child is built by taking from parent A, after the random reorientation, the coordinates of the $N/2$ atoms that are above a $z=z_A$ plane , where $z_A$ is chosen in order to have precisely $N/2$ atoms above it. From parent B, the coordinates of the $N/2$ atoms that are below a suitable $z=z_B$ plane are taken. In case $N$ is odd, the parent with the highest fitness, let's say A, contributes with one extra atom, selected by adjusting the cutting plane $z=z_A$ to the purpose. This procedure produces a list of $N$ lines containing the coordinates. To this list, it is pasted the fixed column of the labels indicating the atomic species. Thus, some of the atoms may swap species.

After assemblage one or more pair of atoms may be found to be ``too close'' (closeness is defined with the same three thresholds mentioned above). This can happen only at the interface between the two pieces of the parents. In this case the two halves are pushed away along the $z$-axis until the minimum distance between pairs is satisfactory. Also this adjustment operation is regarded as {\em mutation} (see below).

This approach is efficient in proposing new structures with the correct composition, but often, due to the interchange of atomic species, it can destroy those local features that determine a high fitness for the parent clusters. We have seen that the use of this crossover operator promotes a rapid decrease in energy of the running GM. However, due to the built-in large variation of local composition at the interface between the halves of the parent clusters, the finding of the actual GM may be hindered.

According to our sampling, also the total spin of the clusters is left free to evolve together with the spatial coordinates of the atoms. In this way we sample on equal footing the configurational space of atomic coordinates and the spin.
The crossover of the spin coordinates is performed in the following way: When we create a new child by grabbing the atomic coordinates from the parents as explained above, we also make note of the atom-projected spin moments (via Hirshfeld partitioning of the electron density) for each atom. Such spin moments are given as initial moments of the individual atoms of the child. During the optimization process, these atom-projected moments are left free to change.

(ii) Crossover-2: This is close to the cut-and-splice crossover operator of Deaven and Ho \cite{crossover}. After the reorientation, atoms with positive $z$-value are selected from one cluster and atoms with negative $z$-value are selected form the other cluster. These complementary fragments are spliced together. In this way the stoichiometry is not necessarily preserved. The choice, here, is to accept the child if the stoichiometry is preserved, otherwise reject it, select new parents, and iterate the procedure until the child has the required stoichiometry \cite{r15,r16}. The advantage of this procedure is that it helps to maintain winning features of the parent molecule but most of the time it takes many attempts to obtain a valid child, even for a moderately sized cluster. In case one or more pairs of atom are too close, we adopt the same remedy as for crossover-1. The spin coordinates are taken care of the same way as in the crossover-1 case.

(iii) Crossover-3: After re-orientation of the selected parent clusters we take all the metal (Mg) atoms from one parent molecule and all the oxygen atoms from the other parent molecule. This crossover helps introducing diversity in the genetic pool, but  the rate of rejection during the assemblage of the child can be rather high due to the high likelihood that two atoms are too close.

Other crossover rules can be designed for this particular system and a totally different one may be needed for a different system. For instance, we have extended our scheme to periodic structures, in order to treat, e.g., structures adsorbed on (not necessarily clean) surfaces and defects in bulk materials. This extension will be presented elsewhere \cite{xunhua}.

The three mentioned schemes are thus not intended to form an exhaustive set of crossover operators, but rather to propose a sufficiently diversified set, so that we can in the following meaningfully test one scheme or combinations of schemes against another in order to find which scheme, or combination, is more robust in finding the GM of the test PES's.

\subsubsection{\bf Mutation}
\label{mut}
After crossover, which generates a child, mutation is introduced.
As for crossover, different mutation operators can be defined. We have adopted a) a rigid translation of the two halves of the clusters relative to each other (this is performed if atoms coming from the two different parents find themselves too close upon splicing of the two halves) and b) exchange of the atom species without perturbing their coordinates (this is included in crossover-1, but not performed after crossover-2 and -3).
We have purposely not introduced any such mutation after crossover-2 and -3. Since we mix different kinds of crossover (as explained in section-\ref{ga-remd}) along with different selection schemes (as discussed in section-\ref{sr} and \ref{ga-remd}), there is always the chance that a species exchange is performed, via crossover-1.

\subsubsection{\bf Similarity of structures}
\label{ss}
In order to decide whether a newly found structure was already seen previously during the GA scan, after the local optimization we a) compare the energy of the new structure with that of all the others seen before and b) use a criterion based on the distances between all the atoms' pairs. In practice, we construct a coarse-grained radial distribution function (rdf) of the clusters, consisting of 14 bins conveniently spaced. Each bin contains the (normalized) number of atom pairs whose mutual distance is included between the two distances that define the boundaries of the bin. For each cluster we have then a 14-dimensional rdf-array and the euclidean distance (i.e., the square root of the sum of the squared difference between corresponding elements in the two arrays) between the arrays arranged for two clusters is evaluated.
If this distance (note that it is a pure number) is greater than a convenient threshold (we used 0.01), then the structures are considered as different \footnote{Note that since the optimizations and the comparisons are performed at different levels of accuracy, from the FF to hybrid DFT functionals, at each step the comparison in energy and geometry is performed against a set of clusters found with the same level of accuracy. In other words we need to store the pool at all levels of accuracy.} In the opposite case, if also the energies of the two clusters are within 0.005 eV, the two structures are considered as {\em similar}.\\
We notice that this very simple similarity criterion uses a descriptor of the cluster, the 14-dimensional rdf-array, that is invariant upon rigid translations and rotations of the cluster, and upon permutation of the atoms of the same species.

\subsubsection{\bf Local optimization and early-rejection scheme}
\label{cascade}
We start by noticing that the child-assemblage step is usually a computationally cheap part of the algorithm (not always, see below), because it requires just some I/O manipulation. The local optimization step is the computationally expensive part of the algorithm, in particular at the {\em ab initio} level, because one needs several energy and force evaluations. For a wide range of clusters, fully {\em ab initio} based global optimization search is computationally expensive, and becomes prohibitive rapidly for larger cluster sizes.
Moreover, if the initial members of the population (randomly generated and then relaxed clusters) are far away from the actual GM, the convergence becomes extremely time-consuming. Therefore, we have adopted a classical FF (namely ReaxFF)\cite{ff,ff1} for performing a computationally inexpensive pre-scan of the PES of the clusters.
\footnote{ReaxFF was chosen because it is a flexible, reactive FF, and its parametrization for Mg and O yields for bulk MgO lattice constant within 1 \% and bulk modulus within 10 \% from the experimental values; similarly, for the MgO and O$_2$ dimer, it gives bond length and vibrational frequency within 1 \% and 5\% from the respective reference values.}

For DFT-based local optimization, we are using the ``trust-radius-method enhanced version of the BFGS optimization algorithm'', as implemented in the full-potential, all-electron numerical-atomic-orbitals-based code FHI-aims,\cite{aims} which is the code we also chose for the evaluation of energies and forces at the DFT level.

In the {\em low-level settings} DFT calculations, forces were converged to less than 0.01 eV/\AA, using the van-der-Waals-corrected \cite{ts-scheme} PBE exchange and correlation functional\cite{pbe}, henceforth labelled PBE+vdW.
The grid settings were set to ``light'' and the basis set to tier-1 \cite{aims}.

The main reason of this intermediate step is the implementation of the {\em early-rejection} scheme. Although, as shown in sections \ref{comp-funct} and \ref{rpa}, the geometry and the energy of the structures is not fully converged with PBE+vdW @ {\em low-level settings}, we have realized that there is a one-to-one mapping between the structures found at this level and those fully converged. In other words, if two structures are {\em similar} (according to the criterion described in section \ref{ss}) with geometries locally optimized with the PBE+vdW @ {\em low-level settings}, then they are {\em similar} also when the geometries are further optimized at the PBE+vdW @ {\em high-level settings} (see below).

In practice, the {\em early-rejection} scheme consists in rejecting those structures that, when optimized with {\em low-level settings}, result in {\em similar} to already known structures or are more than 1.5 eV higher in energy than the current GM. With this (rather conservative) choice, we avoid the risk of rejecting structures that would eventually result in the GM or close to it. Note that child structures that are rejected at this stage because of high energy are not forgotten. For them, the energy at the PBE0+vdW @ {\em high-level settings} is in any case calculated (without further optimization) and their fitness evaluated. Thus, there is a chance that they are selected as parents (in particular in the mixed $\rho_i / \tilde{\rho}_j$ scheme). This PBE0+vdW evaluation is necessary in order to have a comparison with the fully optimized structures using consistent quantities. Furthermore, while passing from PBE to PBE0 can change considerably the relative energies (see below), further optimizing from {\em low-level settings} to {\em high-level settings} lowers the energy by no more than 0.2 eV. Thus the rejected structure enters the competition for the selection with a {\em fitness} evaluated with a large error bar, but still acceptable. On the other hand, it would be unwise to forget such structure, as variety is one necessary ingredient of a successful GA scan.

In the PBE+vdW, {\em high-level settings} optimization, atomic forces were converged to less than $10^{-6}$ eV/\AA. The grid settings where set to ``tight'' and the basis set was tier-2 \cite{aims}.

In cascade, for the structure optimized with PBE+vdW, {\em high-level settings}, we evaluate  (without further optimization) the PBE0+vdW energy with the tier-2 basis set. This energy is later used for the calculation of fitness of that particular cluster.
The difference between binding energy for an isomer optimized with PBE0+vdW forces (tight / tier 1 / forces converged to 10$^{-5}$ eV/\AA) and the binding energy of the same isomer optimized with PBE+vdW (tight / tier 2 / forces converged to 10$^{-5}$ eV/\AA), when the energy of both geometries is evaluated via PBE0+vdW (tight / tier 2), is small, i.e. at most 0.04 eV among all cases we checked. The computational cost of the PBE0+vdW further optimization would be thus not worthy (we estimated a gain of up to a factor 2 of overall computational time just by skipping the latter optimization)

A two-level scheme, reminiscent of our cascade approach, was already introduced in Ref. \onlinecite{r17}, but our approach goes beyond that, e.g., by including the initial pre-screening of structures performed through FF-GA. Furthermore, the fine tuning of the several levels of our cascade
approach (choice of the functionals, of the levels of increasing accuracy) is here motivated by carefully selected benchmark tests, discussed in section \ref{vld}.

\subsubsection{\bf Fixed-spin scheme and perturbative PBE0, based on PBE orbitals}
\label{sec:spin}
An alternative to our scheme, where the spin is left free to evolve during the electronic-structure optimization, is to perform several parallel and independent searches at a given stoichiometry with different fixed spins. In this way, the spin at all levels along the cascade would be fixed, in particular PBE and PBE0 electronic structure calculations (with the same geometry) will have the same spin (as well as the same geometry).
This feature allows for a fascinating possible short-cut for the estimate of the PBE0 (or, in general, of the hybrid xc-functional) energy, i.e., calculating PBE0 xc energy by using PBE orbitals, without performing the self-consistent field (scf) optimization. In practice, this would be a perturbative PBE0 calculation, using the PBE orbitals~\footnote{In the case of PBE0 form PBE orbitals, this evaluation reduces to calculate the exact (Hartree-Fock) exchange for the PBE orbitals and then linearly combine the outcome with PBE exchange, with the ratio $1/4-3/4$} as starting point. Given the perturbative nature of this energy evaluation, the spin of converged PBE and perturbative PBE0 must agree, otherwise the orbitals would be too different. \\
Following this approach, provided that the non-scf PBE0 (relative) energies are a good approximation of the converged PBE0 energies, the advantage in terms of CPU time is evident. However, a) the fact that perturbative PBE0 yields reliable relative energies among the different isomers must be tested for each system and b) the speed up for the single PBE0 calculation is obtained at the expense of the necessity of performing more than one fixed spin GA for each stoichiometry, in fact c) it is not always easy to estimate the full list of possible low-energy spin states for a given stoichiometry. \\
We have tested this alternative route on selected stoichiometric clusters, (namely Mg$_{3}$O$_{3}$, Mg$_{4}$O$_{4}$, Mg$_{10}$O$_{10}$) and non-stoichiometric clusters (namely, MgO$_{4}$, Mg$_{2}$O$_{5}$, Mg$_{4}$O$_{10}$). \\
We have noticed that, perturbative PBE0 yields reasonably good relative energies, compared to the converged PBE0 for some of the stoichiometric cases, namely
Mg$_4$O$_4$ and Mg$_{10}$O$_{10}$.
For these two systems the maximum absolute error by using perturbative PBE0 is 0.25~eV (wrt converged PBE0, over the set of 4 lowest-energy Mg$_4$O$_4$ isomers and 10 lowest-energy Mg$_{10}$O$_{10}$ isomers). This error is evaluated by using as independent reference the energies of the GM in both cases, converged and perturbative, and comparing the energies relative to the respective GM.\\
For Mg$_{3}$O$_{3}$ and the non-stoichiometric clusters, the outcome is rather different. The maximum absolute error is above 1~eV in all those cases and even above 2 eV for all non-stoichiometric cases, where the errors can have both signs.

In order to explain this, we need to mention that stoichiometric clusters in their global minimum are singlets \cite{letter}. For Mg$_{4}$O$_{4}$, Mg$_{10}$O$_{10}$ all the isomers in the energy window we considered where also singlets. Although Mg$_{3}$O$_{3}$'s global minimum structure is a singlet, many of the higher energy isomers admit nearly degenerate spins (singlet and triplet). For non-stoichiometric clusters, we find several global minimum (as well as higher energy) structures with nearly degenerate spins. Upon comparison of PBE and PBE0 Kohn-Sham orbitals for the isomers yielding the largest errors, we attribute the behavior to level crossings, which we see always appearing in connection with quasi-degenerate spin states, for the systems we have examined.

In summary, for those stoichiometric clusters for which no isomers with degenerate spin are found, the fixed spin + perturbative PBE0 is a practical route with computational cost lower than the variable-spin alternative. Therefore, the perturbative PBE0 energies are reliable enough for the GA scheme, in terms of energy-based likelihood of selection of isomers for the crossover. However, accurate converged PBE0 energetics must be evaluated at the end of the GA scan, although only for the low energy clusters.

For all the other cases, the large absolute errors make it impossible to rely on the perturbative-PBE0 energetics for constructing the fitness function.
Furthermore, for the non-stoichiometric clusters, we observe a systematic lower number of iterations for converging the electronic structure when the spin is left free to evolve, compared to the fixed-spin case. In other words, for non-stoichiometric Mg$_M$O$_x$ clusters, the free-spin scf optimization is more efficient that the fixed-spin one.

\subsubsection{\bf Parallelization}
As it should be clear form the analysis of the GA algorithm steps, the operation of selecting from the genetic pool two structures for the mating and the subsequent local optimization of the child, is an operation that can be performed at any moment also when a local optimization of a child is already running.
The algorithm is thus suitable for a very efficient parallelization.


It should be noted here that FHI-aims is very efficiently parallelized\cite{aimsp} and on top of this parallelization we add a second level of parallelization, i.e., we run at the same time several local optimizations, independently. The only communication among such replicas is the selection of the parents that is performed from a common genetic pool. The latter is also updated by each replica at the end of each local optimization.
The local optimizations run independently, i.e., each replica can start a new mating + local optimization cycle right after one is concluded; hence, there is no idling time between cycles.
Thus, we have $n$ local optimizations running in parallel, each requiring $p$ cores \footnote{The number $p$ does not need to be the same for all replicas, but, since we deal with systems of the same size, this is the natural choice. Note that for FF-GA $p=1$}, that fill the $n\times p$ cores required for the algorithm. The scaling behavior is about O($p^{1.5}$) with the number of cores for the local optimization part, as discussed in Ref. \onlinecite{aimsp} . The number $p$ is indeed tuned in order to be sure that the speed-up is still O($p^{1.5}$) for the specific system. The scaling with respect to the $n$ replicas is linear, because the replicas are for the most of the time independent and only at the beginning and at the end of each local optimization, information is shared among the replicas. The first level of parallelization is performed within the FHI-aims code, by means of the MPI environment. The second level is script based: The total $n\times p$ number of cores is divided into $n$ groups, $n$ subdirectories are created and in each of them a cycle of local optimization job runs, each using $p$ cores.

\subsubsection{\bf Global convergence}
A robust criterion for the convergence of a global scanning of a high-dimensional PES does not exist. An operative criterion is to continue the search until, for a ``long while'', no new structure with better fitness than the current optimal one is found. As ``long while'', one can set a time that is several times the time employed to find the current optimal structures. In the present work we scanned for at least double the time needed to find the actual GM.

A GA algorithm applied to structure optimization is thus configured as a sequence of local optimizations, followed by large jumps in the configurational space (the generation of the child by combining two parent structures). The goal of the scanning is to find the GM, but also a large set if not all structures energetically close to the GM. In fact, it is not necessary to know a single GM structure, but also whether other nearly degenerate structures are present. The underlying assumption behind the crossover scheme is that that piecewise the geometrically localized features of the high fit structures (the target of the search) can be randomly formed during the scan (already present in the random initial pool or randomly hit by mutation) and that those ``winning'' local features have a high chance of being propagated, i.e., structures containing those local features have higher fit and have high chance to be selected for generating a new structure. By ``local feature'' we literally mean the local arrangement (e.g., bond distances and angles with first nearest neighbours) of an atom or a group of atoms.

\subsubsection{\bf Mechanical stability of the structures (harmonic analysis)}
\label{ha}
After the DFT-GA search is concluded, we perform a calculation of the harmonic vibrations of each structure within an energy window of 1 eV from the GM. This step has two purposes. The immediate purpose is to identify unstable structures, i.e., structures having imaginary vibrational frequencies. The other purpose is to store the frequencies for the stable structures for the evaluation of the vibrational free energy (see section \ref{sec:aiat}).
In case a structure is found unstable, we perturb its geometry along the unstable mode (i.e., the mode related to the imaginary frequency) and then proceed by locally optimizing it. \footnote{The check of the stability of a structure could be performed also within the GA scan, namely between step 6.e and 6.f. In this case, one would accept stable structures and proceed, while unstable structures would be perturbed along the unstable mode and further geometry optimization would be performed, until a local minimum is found. However, the evaluation of the harmonic frequencies, at the necessary accuracy, is extremely expensive, and it is more convenient to accept some possible saddle points in the pool and perform the stability analysis afterwards.}

Within FHI-aims, harmonic vibrational frequencies are computed via finite differences of the analytic forces.
It should be mentioned here that for MgO clusters we have found many structures that, at the usual recommended level of accuracy (really-tight grid settings, forces converged to $10^{-6}$ eV/\AA), showed soft modes mixed with the rigid-body motions. By inspection, these soft modes were identified as finite rotations of moieties such as O$_2$ with respect to the rest of the clusters. In order to numerically resolve the soft modes, we found that a force convergence criterion at $10^{-7}$ eV/\AA~is needed as well as grid tighter than really-tight. The latter was achieved by setting a radial multiplier \cite{aims} equal to 4 (instead of the default 2) on top of the really-tight grid settings. 


\subsection{Validation of the GA scheme: Replica-Exchange Molecular Dynamics}
\label{vld}
Any GA algorithm that aims at yielding a reliable scanning of the PES, demands for a performance test against an expectedly more reliable scheme. Global searches are hardly exhaustive, but, for the cluster sizes we are currently interested in (less than 100 atoms), we identified Replica-Exchange Molecular Dynamics (REMD)\cite{remd1,remd2,remd3,lmg1} as a reliable tool able to span the configurational space of a given (constant number of particles) system. 

In this approach, various replicas of the system are simulated at the same time, each one at a different temperature, and at intervals the replicas are allowed to exchange complete configurations, according to the importance sampling Monte Carlo acceptance rule, based on the Boltzamnn factors of the involved structures:
\begin{equation}
\textrm{acceptance probability} = \min (1,\exp\left( (\beta_j - \beta_i ) (E_j - E_i) \right)
\end{equation}
where $\beta_{i,j} = 1/k_\textrm{B} T_{i,j}$ and $E_{i,j}$ are the total energies of the two selected structures.
This rule ensures that the sampling remains canonical at all temperatures.\\
In REMD, the high temperature replicas allow the sampling to quickly reach regions in the potential energy surface that would not be accessible in the time scale of a regular low-temperature MD simulation. On the other hand, the low-temperature replicas sample the visited basins with the necessary detail for collecting statistical averages.
This approach is computationally very expensive but capable to provide us the relative ensemble populations of all the isomers at various temperatures. We have implemented REMD with ReaxFF (FF-REMD) as per technical prescriptions discussed in Ref. \onlinecite{remd4}.
Several 24-replicas 100~ns FF-REMD runs in the temperature range $100-1\,200$~K (temperatures follow a geometrical progression, i.e., the {\em ratio} between neighbouring replicas is constant) are performed for different Mg$_M$O$_x$ clusters with varying size. 
We have ensured that after finding the GM, the system was subsequently run for sufficiently long time and no new low-energy structure was found. In practice, the GM was typically found within 10 ns, i.e., the total run was at least 10 times longer than the time needed to find the GM.

Here, the purpose of our REMD sampling is to find the geometrically optimized local minima. Therefore, we have locally optimized structures coming from the low temperature replicas (thereby closer to the local minima), by extracting them at constant (frequent) intervals. Afterwards, we applied the same geometrical criterion for sorting out the similar structures as for cGA and thus constructed a pool of low-energy local minima.

The reason for the reliability granted to this scheme is that REMD samples the system without any bias, exploiting the dynamics of the system i.e., following the integration of the equations of motion. Only if the system is not ergodic (loosely speaking, its configurational space can be partitioned into disconnected regions, such that no trajectory at finite temperature can go from one such region to another), REMD is not able to explore the whole configurational space. The highest temperatures of the REMD sampling were chosen so that the cluster were molten; actually we went to near vaporization and we avoided that occasional vaporization of the clusters scattered the atoms by imposing confining walls \footnote{In practice we imposed a smooth confining potential that is activated only when the distance of one atom from the center of mass of the whole system is larger than a certain relatively large distance, namely 3 times the typical bond distance multiplied by the cubic root of the number of atoms. With the idea that particles stay within a sphere that grows with a volume proportional to the number of atoms}.
The result of the comparison between REMD and our cGA are reported in section \ref{ga-remd}.

\subsection{Validation of the cascade scheme: necessity of DFT-GA after FF-GA}
A legitimate question at this point is whether one really needs to perform a DFT based GA search after the FF based GA search, or whether a DFT post-relaxation of the structures (possibly in a wide energy window) that are found by FF-GA is not enough to find all low-energy structures, as they would be found by a DFT-GA. A related question is whether the pre-screening of structures by FF is not biasing the subsequent DFT-GA search, by restricting the set of structures provided to DFT-GA as starting point to a too limited subset of the configurational space.\\
In order to clarify these issues, we performed two tests. In section \ref{sec:FF/DFT-GA} we show the results for Mg$_2$O$_x$ clusters, i.e., the same system on which we base most of the tutorial sections on phase diagrams.\\
In the first test, the lowest-energy FF-GA structures were relaxed at DFT level and the latter structures were compared to the structures found by DFT-GA for the same system.
In the second test, we compared the structures obtained with FF/DFT-GA (i.e., DFT-GA by starting with structures previously obtained by FF-GA) with DFT-only GA (i.e., DFT-GA by starting from random structures).

\subsection{\textit{Ab initio} Atomistic Thermodynamics}
\label{sec:aiat}
The determination of (meta)stable structures by means of {\em ab initio} thermodynamics has been introduced elsewhere in its original formulation for defects in semiconductors \cite{ms1,ms2,ms3}, adsorption on metal-oxide surfaces \cite{ms4,ms5}, and gas phase clusters \cite{lmg1,Eli}.
Here we briefly review this method, adapting the notation to the case of Mg clusters immersed in an atmosphere composed of gas-phase O$_2$.
This ligand may adsorb onto the clusters:
\begin{equation}
\label{eqn3}
\textrm{Mg}_M + \frac{x}{2}\textrm{O}_2 \leftrightarrows \textrm{Mg}_M\textrm{O}_x
\end{equation}
The equilibrium Mg$_M$O$_x$ compositions will depend on the particular environmental conditions under which a possible experiment on such system is performed. In particular, the equilibrium composition depends on the temperature ($T$) and the chemical potential ($\mu_{\textrm{O}_2}$) of O$_2$, the latter in turn related to the temperature and the partial pressure ($p_{\textrm{O}_{2}}$). The thermodynamically most stable Mg$_M$O$_x$ composition at given ($T,p_{\textrm{O}_{2}}$) is the one which has the lowest formation free energy, $\Delta G_f(T, p_{\textrm{O}_2})$, referred to the pristine cluster Mg$_M$ and the reacting gas O$_2$. The formation free energy can be defined (at a finite temperature in an environment of $O_2$):
\begin{equation}
\label{eqn4}
\Delta G_\textrm{f}(T, p_{\textrm{O}_{2}} ) = F_{\textrm{Mg}_M\textrm{O}_x} (T) - F_{\textrm{Mg}_M} (T) - x\mu_\textrm{O} (T, p_{\textrm{O}_{2}})
\end{equation}
Where $\mu_\textrm{O}=\frac{1}{2}\mu_{\textrm{O}_2}$. Therefore, those structures that minimize Eq. \ref{eqn4} at given ($T, \mu_{\textrm{O}_2}$) will be the preferred ones at those experimental conditions.
The application of \textit{ai}AT to study Mg$_M$O$_x$ clusters proceeds along the following steps: (i) Given a number of Mg atoms $M$, for all oxygen compositions calculate the free energy of the low-energy isomers as found in the GA scan. (ii) Compare the relative thermodynamical stability of those structures ( both stoichiometric and non-stoichiometric compositions), as a function of ($T, p$), i.e., construct a phase diagrams. (iii) Identify the most relevant composition that comes out to be the most stable at a given experimental condition ($T,p$).\cite{lmg1} All the free-energy terms in Eq. \ref{eqn4} can be evaluated at the {\em ab initio} level, below we show in detail how.

In order to evaluate $\Delta G_\textrm{f}(T, p_{\textrm{O}_{2}} )$ from Eq. \ref{eqn4}, one needs to compute the free energies of the cluster+ligands and of the pristine cluster, and the chemical potential of oxygen. For gas-phase clusters, this can be done in terms of their corresponding partition\cite{pf} functions that requires the consideration of translational, rotational, vibrational, electronic, and configurational degrees of freedom.\cite{aiat,lmg1}.
In summary, the free energy of the cluster is calculated as : \footnote{The expression for $F^\textrm{rotational}$ is valid for non-linear structures; for the (few, in our case) linear structures, it 
is modified into: $-k_{\rm B}T\ln(\frac{8 \pi^2 I_A k_\textrm{B}T}{h^2})$ }
\begin{eqnarray}
\nonumber F(T) &=& F^\textrm{translational}(T) + F^\textrm{rotational}(T) + \\
 \nonumber & + & F^\textrm{vibrational}(T) + F^\textrm{symmetry}(T) + \\
\label{eqn16}  & + & F^\textrm{spin}(T) + E^\textrm{DFT}\\
\nonumber F^\textrm{translational} &=& -\frac{3}{2}k_\textrm{B}T\ln \left[ { \left( \frac{2\pi mk_\textrm{B}T}{h^2} \right) } \right] \\
\nonumber F^\textrm{rotational} &=& -k_\textrm{B}T\ln \left[ 8\pi^2 \left( \frac{2\pi k_\textrm{B}T}{h^2} \right)^{\frac{3}{2}} \right] \\
\nonumber &+& \frac{1}{2}k_\textrm{B}T\ln \left( I_A I_B I_C \right) \\
\nonumber F^\textrm{vibrational} &=& \sum_i \frac{h\nu_i}{2} + \sum_i k_\textrm{B}T\ln \left[ 1- \exp \left( -\frac{h\nu_i}{k_\textrm{B}T} \right) \right] \\
\nonumber F^\textrm{symmetry} &=& k_\textrm{B}T\ln\sigma\\
\nonumber F^\textrm{spin} &=& -k_\textrm{B}T\ln \mathcal{M}
\end{eqnarray}
where $E^\textrm{DFT}$ is the DFT total energy, $m$ is the mass, $I_{A,B,C}$ are the three inertia moments of the cluster, the $\nu_i$ are the harmonic vibrational frequencies, $\mathcal{M}$ is the spin multiplicity, and $\sigma$ is the symmetry number. Here, a term $\ln V$ ($V$ is a reference volume), which appears in $F^\textrm{translational}$ is dropped, due to the fact that it cancels out when taking the difference $F_{\textrm{Mg}_M\textrm{O}_x} (T) - F_{\textrm{Mg}_M} (T)$.

The chemical potential of oxygen is calculated as:
\begin{equation}
\begin{split}
\label{eqn13}
\mu_{\textrm{O}_2}(T,p_{\textrm{O}_2}) =
& -k_\textrm{B}T\ln\left[  \left( \frac{2\pi m}{h^2} \right)^{\frac{3}{2}} \left( k_\textrm{B}T \right)^{\frac{5}{2}} \right]\\
& +k_\textrm{B}T\ln p_{\textrm{O}_2} -k_\textrm{B}T\ln(\frac{8 \pi^2 I_A k_\textrm{B}T}{h^2}) \\
& +\frac{h\nu_\textrm{OO}}{2}+k_\textrm{B}T\ln \left[ 1- \exp \left( -\frac{h\nu_\textrm{OO}}{k_\textrm{B}T} \right) \right]\\
& +E^\textrm{DFT}-k_\textrm{B}T\ln \mathcal{M} +k_\textrm{B}T\ln \sigma
\end{split}
\end{equation}
where $\nu_\textrm{OO}$ is the O-O stretching frequency.
The latter relation allows for the one-to-one mapping of the reactant's (oxygen) chemical potential to its partial pressure.

\section{Results}
\subsection{FF-GA vs FF-REMD simulation: Validation of (FF-)GA scanning}
\label{ga-remd}

\begin{figure}[h!]
\includegraphics[width=0.55\textwidth,clip]{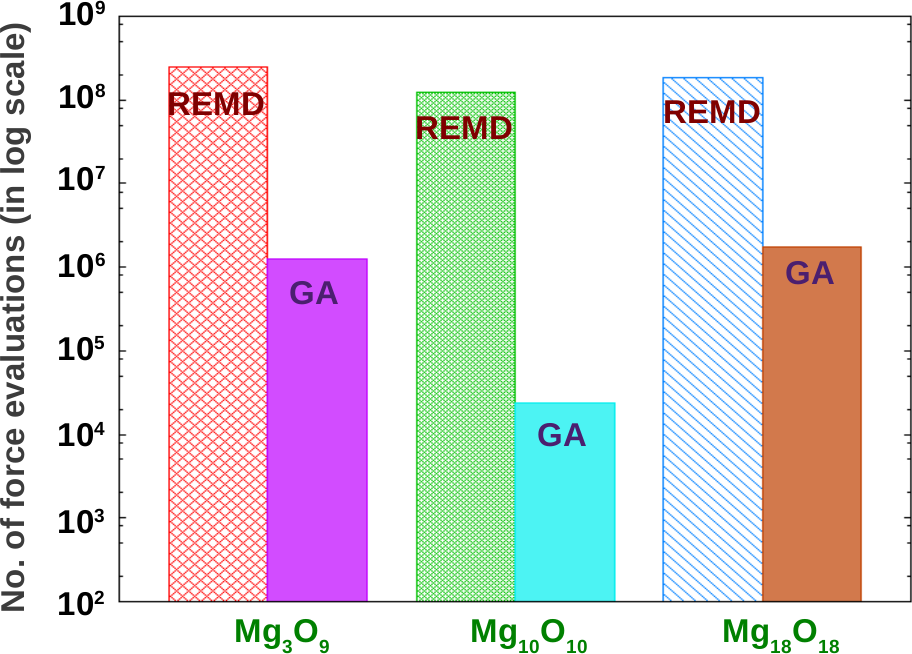}
\caption{Comparison of performance of FF-GA and FF-REMD to locate the same global minima. FF-GA takes orders of magnitude less number of force evaluations than FF-REMD.}
\label{remd}
\end{figure}

In Fig. \ref{remd} we plot the total number of force evaluations needed by two different global PES scanning methods (namely, FF-REMD and FF-GA, respectively) to locate the same global minimum. Unsurprisingly, we see in Fig. \ref{remd} that to find the GM, GA takes a significantly smaller number of force evaluations than REMD \footnote{REMD gives much more information than the local minima \cite{lmg1}. This information was not used here, as we were interested only in the reliability of the adopted cGA scheme.}.
It is important to mention here that the average number of force evaluations to get one minimum using FF-REMD could depend on the provided initial random structures. In this case we have given the same initial structures to both GA and REMD. In order to alleviate the effect of the initial random structures on REMD output, we have tried three completely different set of initial random structures and performed 100ns REMD run each. What we show in Fig. \ref{remd} is the average over these three initial sets. Note however, that the order of magnitude of the number of force evaluations (i.e., MD steps) needed to find the global minimum was found to not change with the initial conditions.

\begin{figure}[t!]
\includegraphics[width=0.55\textwidth,clip]{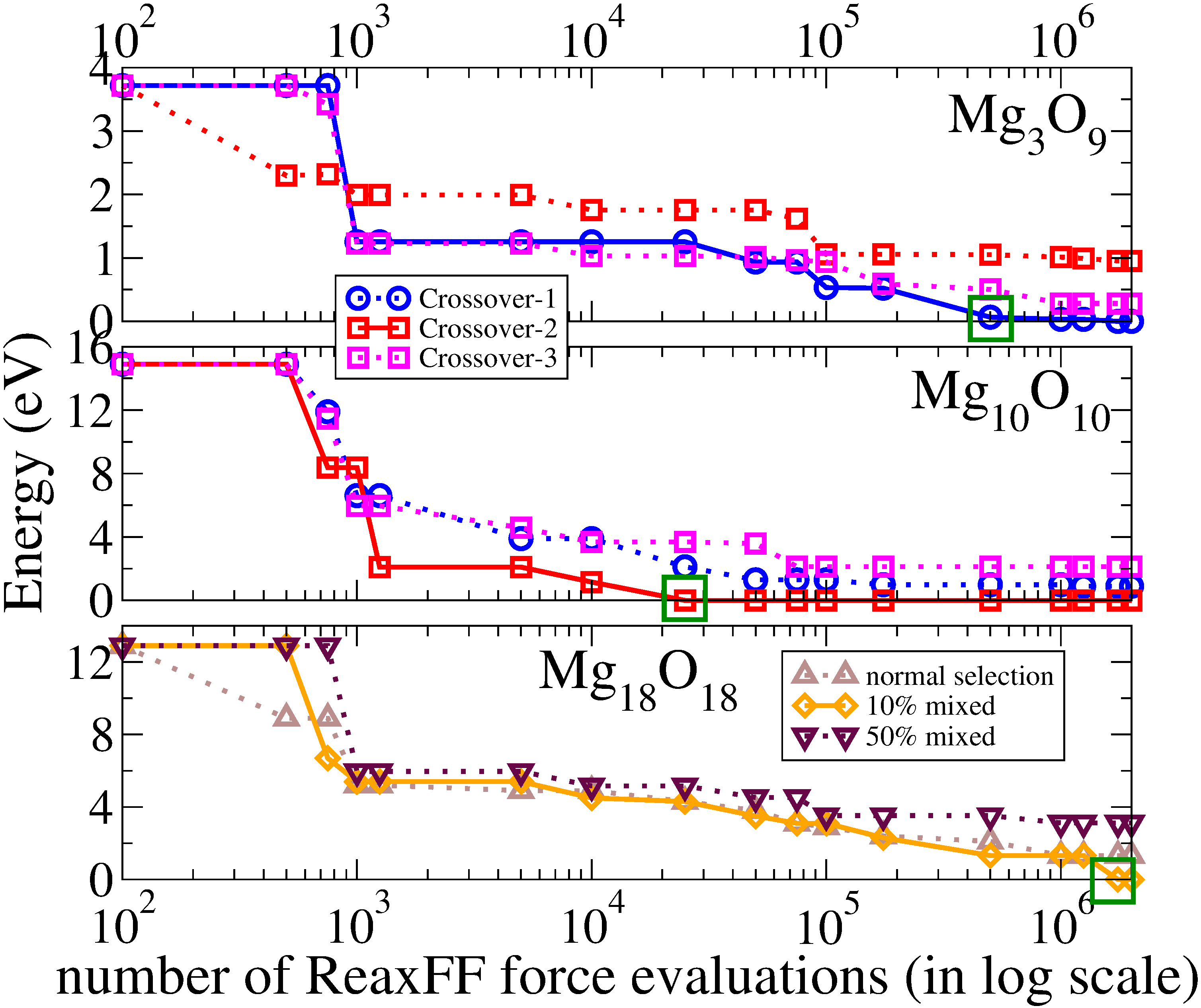}
\caption{Comparison of performance of FF-GA using different crossover schemes and selection criteria. For a given number of FF force evaluations, the difference between the energy of the running lowest-energy structure and the actual GM (as found by a 100 ns REMD simulation) is reported. The green box mark is the number of force evaluations at which the GM is found by the schemes that actually find it.}
\label{ff-ga}
\end{figure}

We have compared the performance of different FF-GA schemes among themselves and against REMD. For all the FF-GA schemes, we set a maximum number of force evaluations to $2\times10^{6}$.
We have found that for systems like Mg$_3$O$_9$, i.e., with very high O-coverage, the GM is found using crossover-1 (as discussed in methods section), but not with crossover-2 and -3 alone, as shown in the top two panles of Fig. \ref{ff-ga}. For these plots, we ran FF-GA for $2\times10^6$ force evaluations and we checked whether the GM was indeed found by comparing to 100 ns long REMD scan.
In some other cases (e.g., Mg$_{10}$O$_{10}$), this situation is changed. Here, crossover-2 could find the global minimum, while crossover-1 and -3 could not. Of course, when we say that the GM is not found by some crossover operator, it means that it is not found within the defined number of force evaluations. It may well be found over (much) longer times. However, what we point out is that one of the schemes finds the GM within the given time, while the other(s) do not, over a time that is at least twice longer than the time required by the scheme that finds the GM to find it. The lesson we learn from this comparisons is that one should then combine crossover operators in order to enhance the chance to find the GM.

\begin{figure}[t!]
\includegraphics[width=0.55\textwidth,clip]{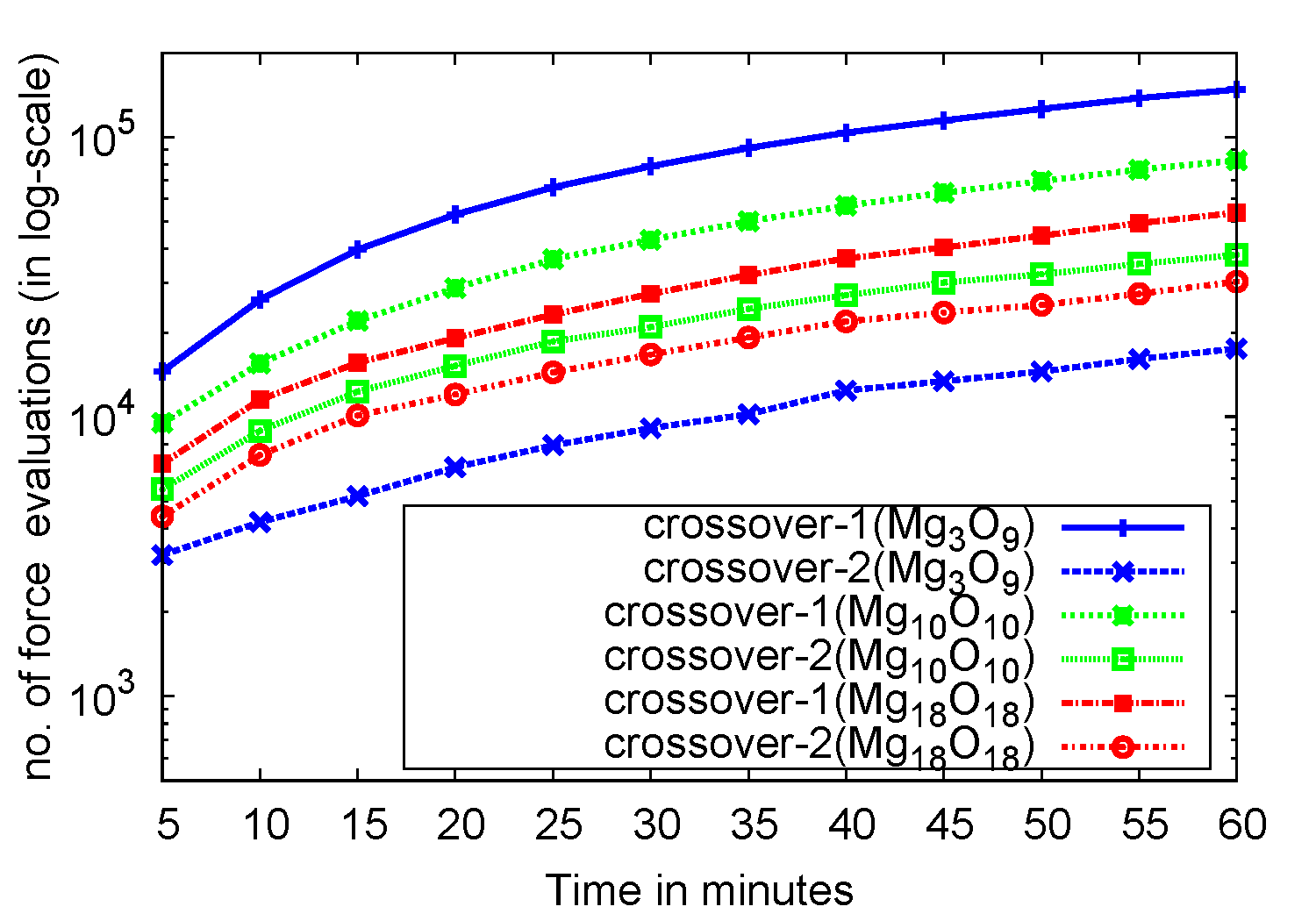}
\caption{Comparison of FF-GA with crossover-1 vs crossover-2. Shown is the number of force evaluations performed after a given CPU time on the same machine.}
\label{cross}
\end{figure}

We note that when designing the GA scheme, one needs to realize that also the average time spent in FF-GA for generating new candidate structures is not the same for different crossover schemes. For instance, in crossover-1 there is no rejection of candidate structures, because the stoichiometry is forced by construction, while crossover-2 may use several trials before hitting the right stoichiometry.
The additional time required to obtain a successful candidate using crossover-2 becomes increasingly large with cluster size, in particular for very unbalanced stoichiometries, i.e., number of oxygen (much) larger or smaller than the number of magnesium atoms. In the case of FF-GA, the time spent in creating a successful candidate with crossover-2 becomes so large that it is even comparable to the time spent in the subsequent local optimization. In Fig. \ref{cross}, we have compared the number of force evaluations performed by crossover-1 and crossover-2 after a given CPU time, for a FF-GA scan. It is clear that crossover-2 always performs less number of force evaluations, irrespective of size of the cluster. The additional time needed by FF-GA using crossover-2 is due to many rejection moves when the correct stoichiometry is not matched after performing the cut-and-splice operation. In the case of DFT-GA this problem is less important, since force evaluations are orders of magnitude more expensive than for FF-GA. Since however any single replica in DFT-GA is performed on several CPUs, if the crossover move is performed on a single CPU then also in this case the time spent in blindly producing the right stoichiometry may be noticeable.

The choice of the crossover scheme is not the only tuning parameter that determines the efficiency of the GA scheme. Another crucial parameter is the selection criterion. If the parent clusters are selected for the crossover with probability proportional to their fitness function, and many similar structures populate the high fitness (low energy) region of the genetic pool, then the selection always fishes among very similar structures and escaping from the basin may be very difficult, also with different mutation schemes. The situation in which there are many similar structures with similar energy is not unusual for clusters. In facts, we have observed it in many cases. We have investigated the benefits of adding the possibility for selecting low fitness function structures, in order to ensure more diversity in the crossover scheme. We show the results for the case of Mg$_{18}$O$_{18}$. In practice, since the fitness function has values between 0 and 1, we have defined a complementary fitness function, $\tilde{\rho}_i$, such that the higher-energy structures have high likelihood to be selected, and vice versa for the low-energy structures. We have tried a scheme with 50\% likelihood to select two structures according to the fitness function $\rho_i$, and 50\% probability that one structure $i$ is selected by using the usual fitness function $\rho_i$ and the other structure $j$ by using the complementary fitness function $\tilde{\rho}_j = 1-\rho_j$. Another scheme uses 10\% of the  $\rho_i / \tilde{\rho}_j$ selection criterion and 90\% of the usual $\rho_i / \rho_j$ criterion. These two variants are compared to the usual scheme (100\% the $\rho_i / \rho_j$ criterion), using REMD as reference for the GM. The results are shown in Fig. \ref{ff-ga}, bottom panel.
We find that, while neither the normal nor the 50\%-mixing scheme is able to find the global minimum in the allotted time, the 10\% scheme eventually finds the global minimum as predicted by REMD. We conclude that it is beneficial that a small percent of high energy structures are used to form new candidate structures.


\subsection{\textit{Ab initio} GA and DFT functionals}
\label{comp-funct}

\begin{figure} [h!]
\includegraphics[width=0.47\textwidth,clip]{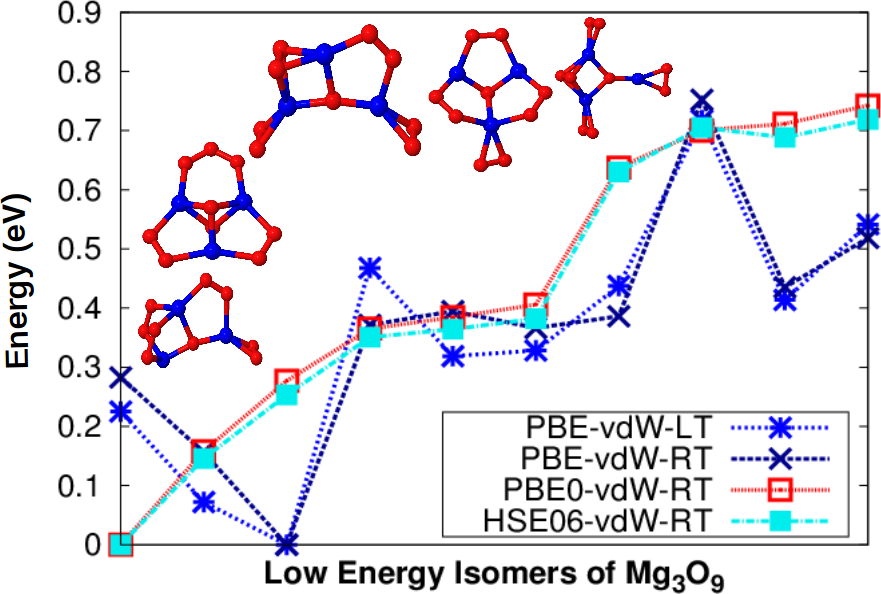}
\includegraphics[width=0.47\textwidth,clip]{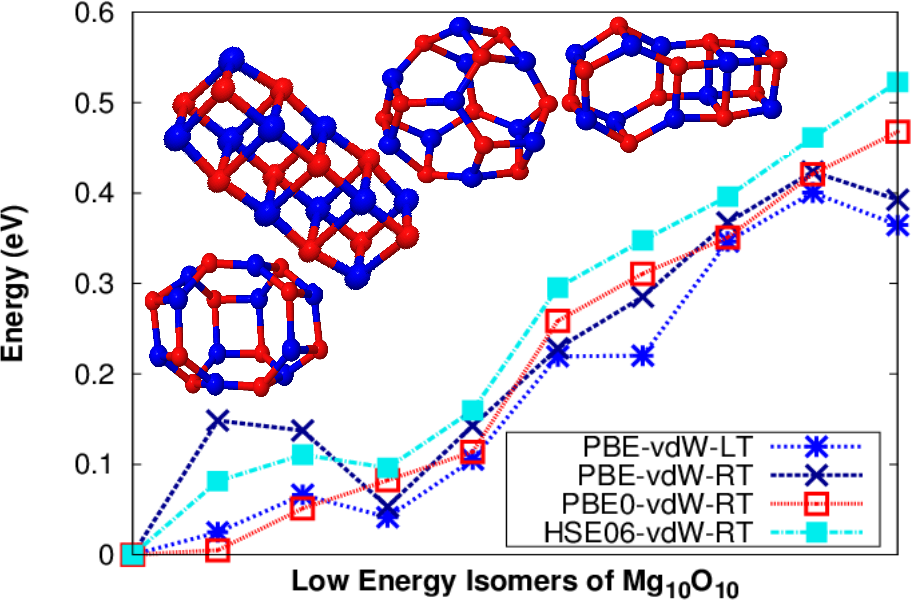}
\caption{Energy hierarchy of same low-energy isomers of Mg$_3$O$_{9}$ (left) and Mg$_{10}$O$_{10}$ (right), using different functionals and settings. The five structures shown are those of the five first points from the left. ``LT'' stays for light grid settings and ``RT'' for really tight. Red and blue spheres represent O and Mg, respectively.}
\label{Mg10O10}
\end{figure}

As already mentioned in section \ref{pgs}, we performed a cascade algorithm, using increasing level of accuracy.
The preliminary scan is performed using FF-GA.  The set of lowest binding-energy structures, within 4 eV from the GM, obtained from FF-GA is used
as initial pool for a DFT-based GA structure search. To give a flavor, this threshold selects for the DFT-GA initial pool 20 structures
for a small system like MgO$_4$ and 300 for the largest system we report here, Mg$_{10}$O$_{10}$. In practice, the selected structures are at first optimized at the DFT level.
In some cases (e.g., for Mg$_{10}$O$_{10}$), this operation already yields the actual GM, as confirmed by our DFT-GA and also found in other studies for stoichiometric clusters.\cite{r6,r7} But, in most of the cases the lowest energy of such structures turned out to be more than 2 eV higher in energy than the actual global minimum.

In Fig. \ref{Mg10O10} we elucidate the reason why we need to go down to the highest level of accuracy for the evaluation of the fitness.
In these figures, we show the relative energy of the lowest-energy isomers of Mg$_3$O$_{9}$ and Mg$_{10}$O$_{10}$, calculated with various xc functionals (we have included another hybrid functional, HSE06+vdW \cite{hse06}, for highlighting the need to go at the hybrid-functional level in the density-functional hierarchy, in order to have converged energetics). The ordering of the structures on the $x$-axis is such that the energies are monotonically increasing for PBE0+vdW (i.e., the functional onto which we base the actual energy hierarchy in our GA scan).

The energy hierarchy predicted for Mg$_3$O$_{9}$ and Mg$_{10}$O$_{10}$ clusters depends significantly on the settings and on the employed functional. In a global search method such as GA, one of the key issues is to accurately calculate the total energies, because the probability of a parent selection fully depends on it. On the other hand, it would be unwise to do the whole optimization only at the highest level of accuracy.
In general, a GA scheme spends most of the time in optimizing structures that eventually turn out to be at high energy (or have been already seen).
Although the energy hierarchy is quite different with different methods, we have checked that very high-energy structures (i.e., more than one 1 eV form the global minimum) are recognized as such already at the lowest-accuracy DFT level. Furthermore, structures that are recognized as not seen at the lowest-accuracy DFT level almost always turn out to be new structures also at the highest-accuracy level. In this sense the lower-accuracy steps in our approach provide a formidable pre-screening of structures, saving a significant amount of computational time.

Here we underline that our GA performs the optimization in the coordinate and spin space at the same time. We find that while stoichiometric clusters are always singlet in the ground state, non-stoichiometric clusters exist typically in different spin states, almost degenerate. This delicate issue is the object of a different publication \cite{letter}.

\begin{figure}[h!]
\centering
\includegraphics[width=0.5\textwidth,clip]{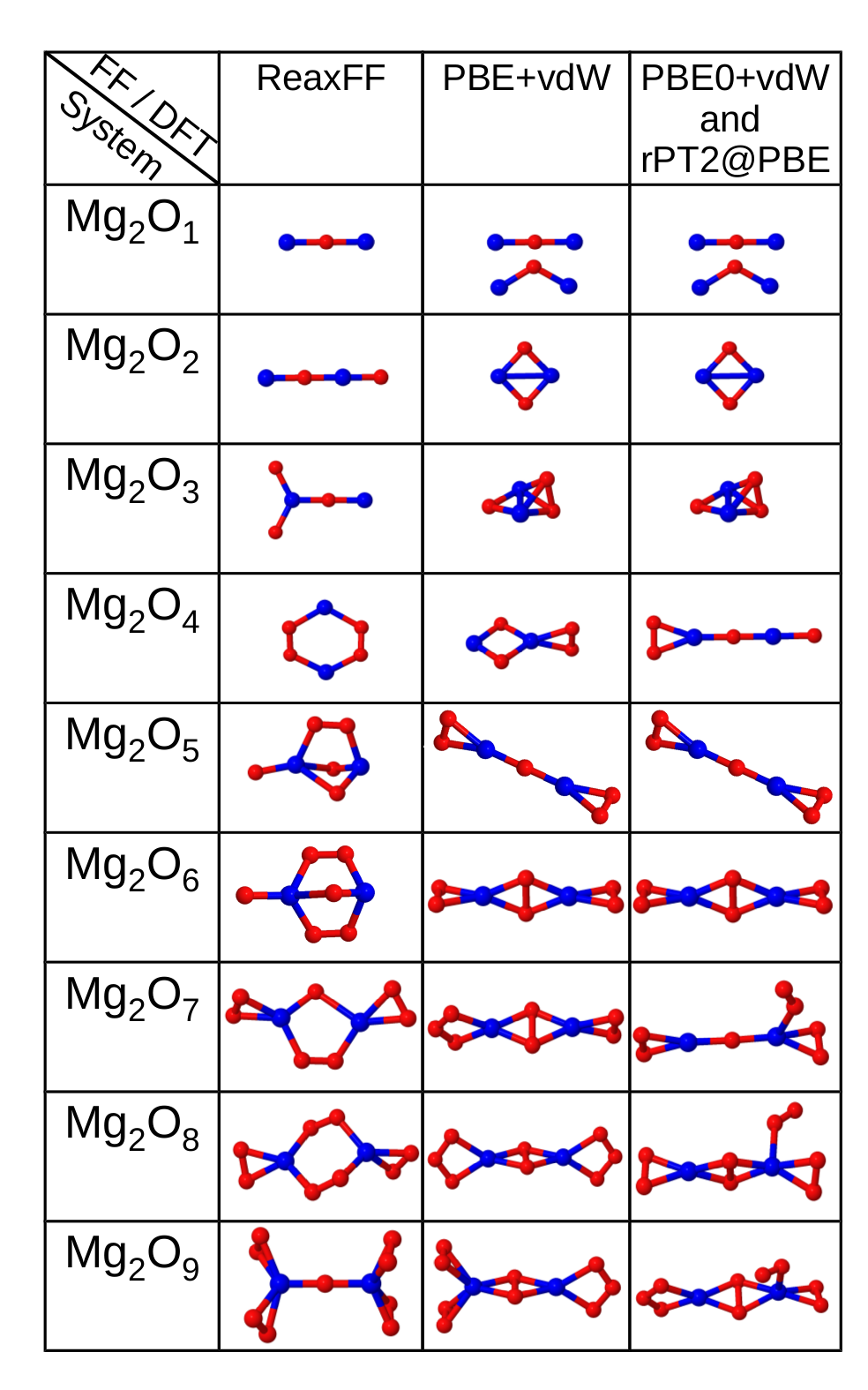}
\caption{Comparison of the structure of the lowest energy isomer for Mg$_2$O$_x$ at each coverage, as found by reaxFF, PBE+vdW, PBE0+vdW, and rPT2@PBE (the latter two always give the same energy hierarchy for the shown compositions, hence the use of only one shared column in the table). The geometries for PBE0+vdW, and rPT2@PBE are relaxed with PBE+vdW and only the energies are evaluated with the two higher level methods. Red and blue spheres represent O and Mg, respectively.}
\label{FT:Mg2Ox-FF-PBE-RPA}
\end{figure}

\subsection{Effect of xc-functional on formation energy of MgO$_x$.}
\label{rpa}

The accuracy of PBE0+vdW energies was validated against the highest levels currently achievable within the DFT framework: Exact exchange plus correlation in the random-phase approximation (RPA) with renormalized single-excitations correction (rSE) plus second-order screened exchange (SOSEX). RPA+rSE+SOSEX is collectively referred as renormalized second-order perturbation theory, rPT2. The corrections are applied on PBE and PBE0 orbitals (the resulting functionals are thus indicated as RPA+rSE@PBE, RPA+rSE@PBE0, rPT2@PBE, and rPT2@PBE0~\cite{xinguo2,xinguo3}, and collectively as beyond-RPA in the rest of the text).~\footnote{RPA+rSE@PBE, RPA+rSE@PBE0, rPT2@PBE, and rPT2@PBE0 energies were calculated with FHI-aims, using ``really tight" grid settings and ``tier 4" basis set. Energies for all these high-level functionals are counterpoise corrected for the basis set superposition error}.

The differences between the different levels of theory are then evident also in the comparison of the cluster structures, as shown in Fig. \ref{FT:Mg2Ox-FF-PBE-RPA}. The geometries are relaxed with PBE+vdW, and only the energies are evaluated with PBE0+vdW and beyond-RPA methods. Nonetheless, the hierarchy of the isomer changes, when passing from PBE+vdW to PBE0+vdW and beyond-RPA, in particular at high coverages.
We can now look in closer detail into the comparison among different functionals.

\begin{figure}
\includegraphics[width=0.47\textwidth,clip]{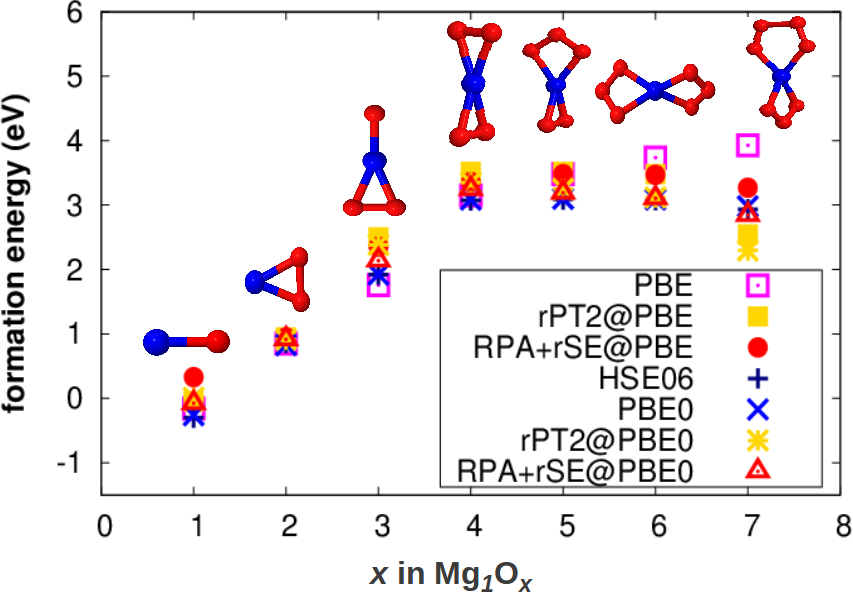}
\includegraphics[width=0.47\textwidth,clip]{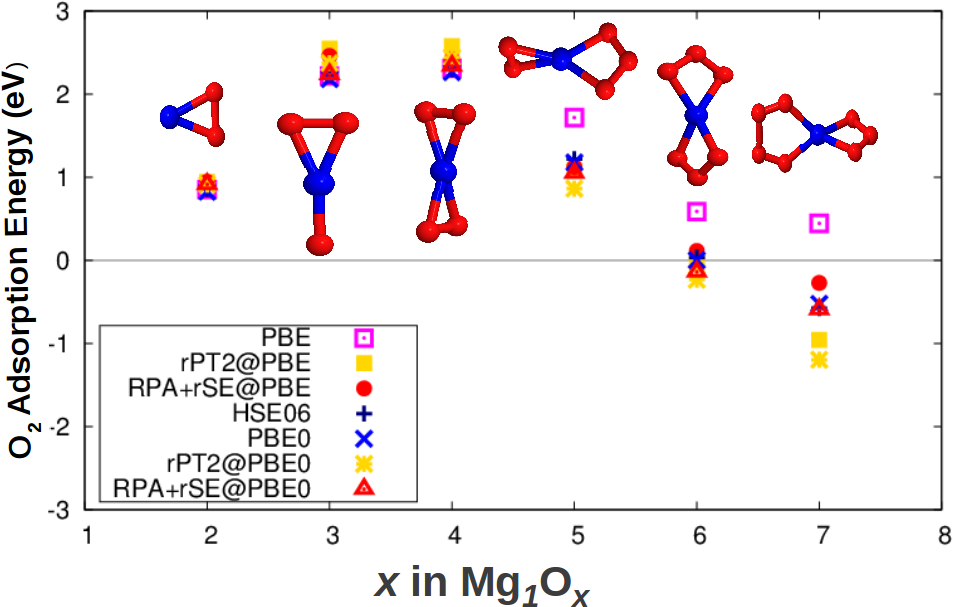}
\caption{Formation energy of MgO$_x$ clusters (left) and energy of O$_2$ adsorption on MgO$_x$ clusters (right), using different xc-functionals. Red and blue spheres represent O and Mg, respectively.}
\label{be1}
\end{figure}

We have calculated the formation energy of MgO$_x$ (for $x=1,2,...,7$, see Fig. \ref{be1}, left panel) and the adsorption energy of O$_2$ (see Fig. \ref{be1}, right panel) on such clusters. We compared the results using different functionals (namely PBE+vdW, PBE0+vdW, HSE06+vdW, beyond-RPA). The formation energy ($\Delta E^\textrm{form}$) (i.e., atomization energy) of a cluster (e.g., MgO$_x$) is defined as:
\begin{equation}
\Delta E^\textrm{form} = E(\textrm{Mg}) + \frac{x}{2}E(\textrm{O}_2) - E(\textrm{MgO}_x)
\end{equation}
where $E(\textrm{MgO}_x)$, $E(\textrm{Mg})$ and $E(\textrm{O}_2)$ are the total energies of MgO$_x$, Mg and O$_2$, respectively. At each stoichiometry, the same geometry was used ofr all functional, namely the PBE0+vdW GM.

Fig. \ref{be1} (right panel) shows how the adsorption energy of a single O$_2$ molecule changes with changing stoichiometry of MgO$_x$ clusters. The adsorption energy ($\Delta E^\textrm{ads}$) of O$_2$ is defined as:
\begin{equation}
\Delta E^\textrm{ads} = E(\textrm{MgO}_{x-2}) + E(\textrm{O}_2)  - E(\textrm{MgO}_x)
\end{equation}
where for both MgO$_x$ and MgO$_{x-2}$ the GM structures were used.

We find that  PBE+vdW strongly overestimates the adsorption energy at larger $x$, resulting in a qualitatively incorrect prediction that O$_2$ adsorption would be favored over desorption up to a large excess of oxygen. Such behavior is not confirmed by hybrid functionals or beyond-RPA. The reason for this failure of PBE+vdW is the self-interaction error, which favors electron sharing among a large number of anions.

\subsection{FF-GA with DFT post-relaxation vs DFT-GA starting from FF-GA structures}
\label{sec:FF/DFT-GA}
\begin{figure}
\centering
\includegraphics[width=0.55\textwidth,clip]{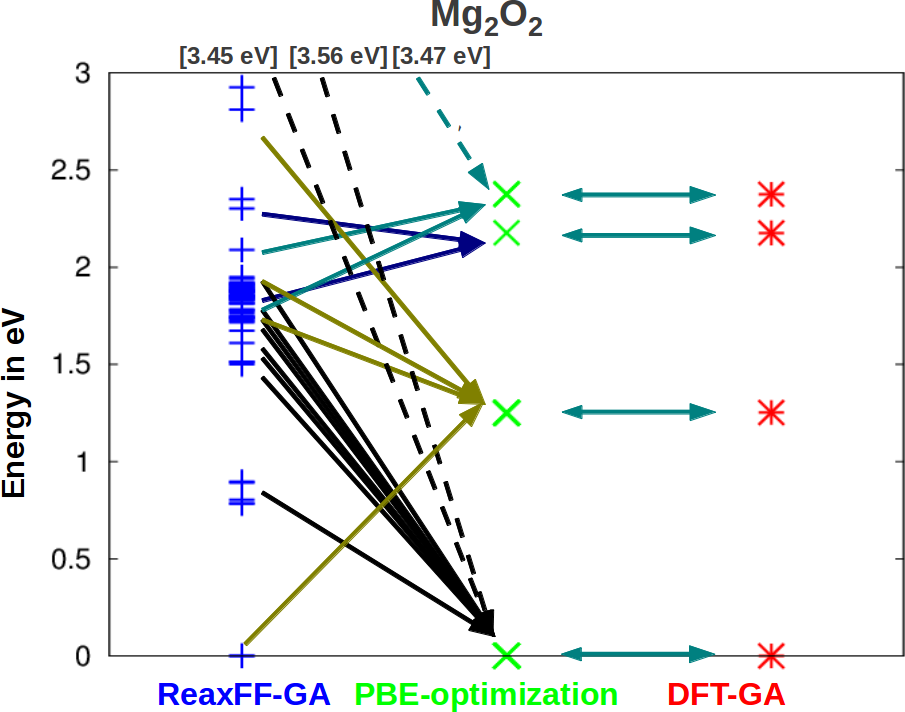}\\
\caption{Comparison, for the stoichiometric MG$_2$O$_2$ cluster composition, of isomers obtained by FF-GA (left column), subsequently relaxed via DFT (center column), and isomers obtained by a direct DFT-GA search started from random structures (right column). The clusters are identified by their relative binding energy and the zero in the energy scale is the GM for each method. The single-side arrows connect the FF structures to their corresponding DFT-relaxed structures, while the double-side arrows connect the same structures found by DFT via the two different routes. The dashed arrows departing from FF-GM pointing to an energy value show at what energy, relative to DFT-GM, relaxes via DFT the FF-GM. The energy values on the top horizontal axis show the energy (relative to FF-GM) of the FF-isomers that relax to the DFT-isomers inside the plot and linked via the downward dashed arrows.}
\label{FF-DFT-GA-1}
\end{figure}

\begin{figure}
\centering
\includegraphics[width=0.55\textwidth,clip]{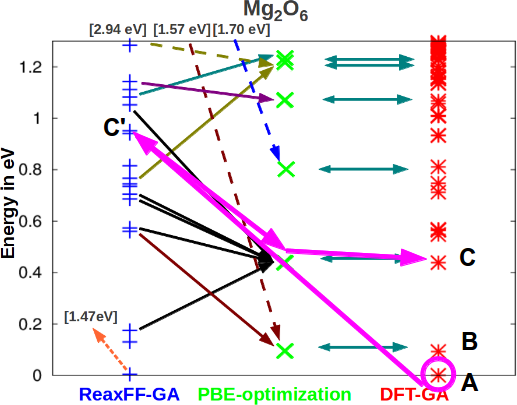}\\
\vspace*{12pt}
\includegraphics[width=0.3\textwidth,clip]{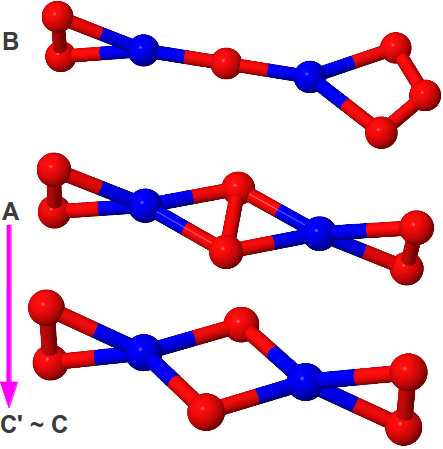}
\caption{Top panel: Comparison for the non-stoichiometric MG$_2$O$_6$ cluster compositions, of isomers obtained by FF-GA (left column), subsequently relaxed via DFT (center column), and isomers obtained by a direct DFT-GA search started from random structures (right column). The symbols and thin arrows have the same meaning as in Fig. \ref{FF-DFT-GA-1}. The purple thick arrows link the optimzation pathway from isomer A (optimized via DFT) through isomer C' (optimized via ReaxFF by starting from A), to isomer C (isomer C optimized via DFT). In the bottom panel, the structures of isomers A and C are shown together with isomer B, i.e. the second-best isomer according to DFT energy hierarchy. In this representation, isomer C' would be undistinguishable from isomer C.}
\label{FF-DFT-GA-2}
\end{figure}

In Figs. \ref{FF-DFT-GA-1} and \ref{FF-DFT-GA-2} we show, for Mg$_2$O$_2$ viz. Mg$_2$O$_6$, the lowest-energy FF-GA structures (left column), the same structures after relaxation at the DFT level (center column), and the lowest energy DFT-GA structures (right column). The isomers are identified by their energy. The one-side arrows, from FF-GA to DFT-relaxed structures, connect initial to final point of the DFT relaxation. The two-sided arrows connect the same structures found via two different routes. The energy window was chosen in order to include around 100 FF-GA structures.
FF clusters (left column) that are not linked to DFT clusters relax to structures that are outside the energy window. The DFT structures (right column) that are not linked to the center column can be a) structures obtained from FF structures outside the shown energy window or b) structures that {\em cannot} be found by using the chosen FF for the pre-screening (see below).

It is immediate to note that a) often, several FF structures relax to the same DFT structure, and b) most, but crucially not all the DFT local minima can be obtained by direct local relaxation of FF local minima, even though with in general very different energetic ordering between reaxFF and DFT. In particular, reaxFF finds all the low-energy DFT structures for Mg$_2$O$_2$ but misses the DFT-GM for Mg$_2$O$_6$.

This is a remarkable failure of a strategy which post-optimizes via DFT a set of FF structures. We further analyze this aspect with the aid of the bottom panel in Fig. \ref{FF-DFT-GA-2}. The GM as found by DFT (structure A) is quite different from the second most stable in (DFT) energy hierarchy (B). The latter is also found by optimizing some of the low-energy FF structures. In order to show that structure A {\em cannot} be found by post-optimizing via DFT FF-optimized structures, we have performed the following test. Isomer A was optimized via FF. This gave isomer C'. The latter was then optimized via DFT. This gave isomer C, which is indistinguishable from C' upon visual inspection, but rather different from A. We conclude that the FF used in the example is totally blind to Mg$_2$O$_6$ DFT-GM. Note that for those ``DFT-GA'' structures that are also found by minimizing ``reaxFF-GA'' structures, the DFT$\rightarrow$FF$\rightarrow$DFT optimization path indeed leads back to the starting DFT structure.

Furthermore, we compared the structures obtained with FF/DFT-GA with DFT-only GA for the same systems. The structures obtained with the two approaches were found to be identical within the considered energy window. \\
The results of these tests indicate that reaxFF does not introduce a strong bias that would affect results of FF/DFT-GA. However, just locally relaxing structures found via FF-GA with DFT would not be sufficient for an accurate sampling of the DFT PES, especially at large O$_2$ coverages.\\
The tests were performed also at other sizes, but all the main features are shown in the presented example. In particular reaxFF is in general blind to the DFT-GM (in the sense clarified above) for non-stoichiometric structures. This example is strictly valid only for reaxFF, so one could think in principle that a way to avoid DFT-GA is to use a better potential or to fit a potential on-the-fly with a training-set based on the DFT structures already found.
The caveat we put forward, though, is that for clusters and in particular small clusters, a classical potential may be simply never be able to capture with one set of parameters all the subtle features displayed by high-level DFT xc functionals.
The valuable merit of a FF-GA used as a pre-screening is to ease the building of an initial set of structures for DFT-GA.

\subsection{Phase diagrams via {\em ai}AT: A case study for Mg$_2$O$_x$}

\begin{figure}
\includegraphics[width=0.47\textwidth,clip]{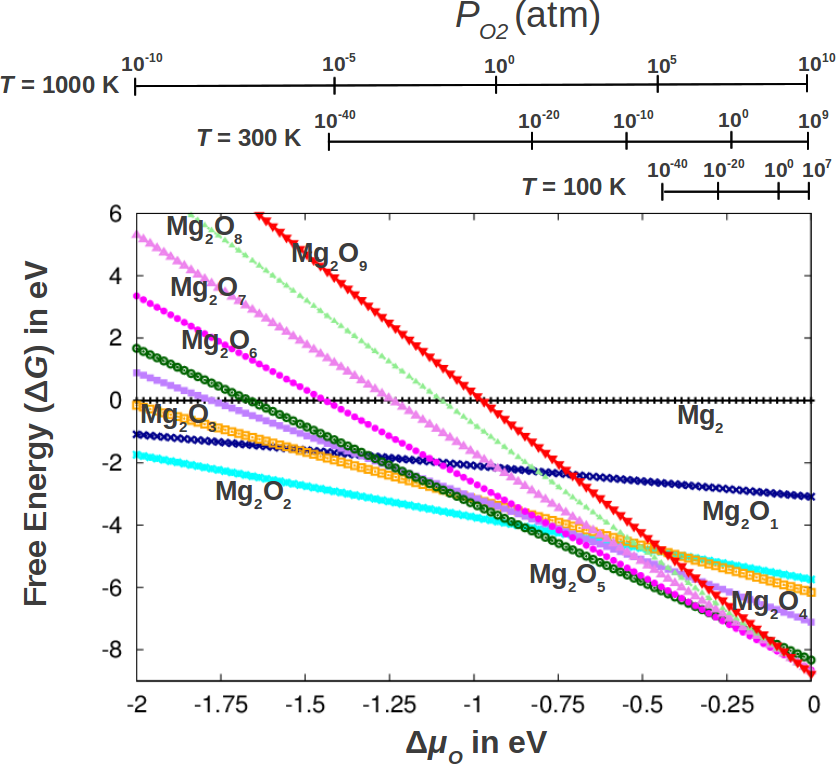}
\includegraphics[width=0.47\textwidth,clip]{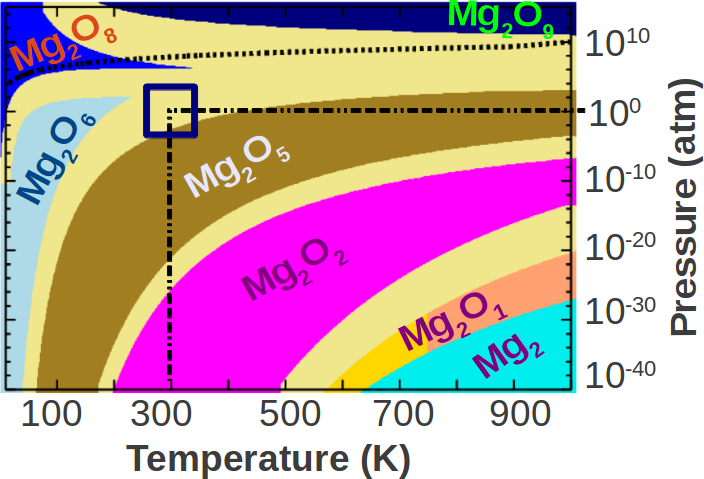}
\caption{Left panel: Formation free energy of Mg$_2$O$_x$ cluster with varying chemical potential of O$_2$. The free energy of the Mg$_2$O$_x$ and Mg$_2$ (see Eq. \ref{eqn4} and \ref{eqn16}) is approximated by their $E^\textrm{DFT}$. Also shown are the pressure scales at $T=100$, 300, and 1000 K.
Right panel: $(p_{\textrm{O}_2},T)$ phase diagram of Mg$_2$O$_x$ cluster. The geometries are optimized with PBE+vdW and the electronic energy are calculated using PBE0+vdW. The sand-colored unlabeled regions are regions where different compositions (at least the adjacent ones) coexist. For Mg$_2$O in the region on the left of its domain two isomers coexist, while in the region on the right only one isomer is found , hence the use of two different colors. The criterion for coexistence is that the free energy of the competing compositions/structures are within 3 $k_\textrm{B}T$ (see text). The square encompasses the region around normal conditions ($T$ = 300~K, $p_{\textrm{O}_2} = 1$~atm) and the dashed-dotted lines are guides for the eyes for identifying the point at normal conditions on the diagram.}
\label{pt2}
\end{figure}

For tutorial purposes, in Fig. \ref{pt2} (left panel), we have plotted the formation free energy $\Delta G$ of a selected cluster+ligand system ( Mg$_2$O$_x$) with varying chemical potential of O$_2$ at a finite temperature ($T=300$ K). The free energy of the Mg$_2$O$_x$ and Mg$_2$ (see Eq. \ref{eqn4} and \ref{eqn16}) is approximated by their $E^\textrm{DFT}$. In this way \cite{ms3,ms4} the only pressure and temperature dependence in the formation free energy is given by $\mu_\textrm{O}$. Therefore, the mapping between chemical potential and pressure at different temperatures differ only by factor in the pressure (log) scale. The pressure axes are calculated according to the relation between $\mu_{\textrm{O}}$ with $p_{\textrm{O}_2}$ as shown in Eq. \ref{eqn13} and \ref{Eq:deltamu}.

We see that at very low pressure at all temperatures the Mg$_2$O$_2$ conformer is the most stable. At $T=300$ K, in a realistically accessible pressure range (i.e., $10^{-15}-10$ atm) the Mg$_2$O$_5$ conformer is the most stable phase. As we move further, at $T=300$ K, in the region at very high pressure, Mg$_2$O$_6$, Mg$_2$O$_8$, and finally Mg$_2$O$_9$ become more favorable.

The temperature and pressure dependence can be combined in a phase diagram as a function of both $T$ and $p_{\textrm{O}_2}$. This is shown again for Mg$_2$O$_x$ in Fig. \ref{pt2} (right panel) where at each $T$ and $p_{\textrm{O}_2}$ a color signals that a particular composition (or combination of compositions, see below) is free-energetically more stable. In this case the fully-fledged free energy of Mg$_2$O$_x$ and Mg$_2$ is considered.

Maybe the most striking aspect of this phase diagram is that the stoichiometric phases (namely, Mg$_2$O$_2$) are predicted to be thermodynamically less stable than non-stoichiometric phases (e.g., Mg$_2$O$_5$, Mg$_2$O$_6$ etc.) at experimentally relevant environmental conditions (e.g., 1 Pa $\le$ $p_{\textrm{O}_2} \le 1$~atm, room temperature (300K)). This finding, which is valid also at sizes different from Mg$_2$O$_x$, is the subject of a different publication \cite{letter}.

In Appendix we show the comparisons of the $(T, p_{\textrm{O}_2}) $ phase diagrams of Mg$_2$O$_x$, obtained at different levels of theory (reaxFF, PBE+vdW, PBE0+vdW, PBE0 without vdW, and beyond-RPA) and by switching on and off different entropic contributions to the free energy (translational, rotational, and vibrational, according to Eq. \ref{eqn16}). Furthermore, we show the effect of the (in)accuracies of O$_2$-binding-energy according to several functionals. In summary, we find that a) PBE+vdW functional is not sufficient even for a qualitative description of the phase diagram of these systems, whereas PBE0+vdW gives qualitatively and quantitatively performances similar to beyond-RPA, b) neglecting all entropic terms, or including only some of those, results only in slight changes in the phase diagrams, comparable to the differences between PBE0+vdW and beyond-RPA diagrams, c) correction of the O$_2$ binding energy error increases the difference between PBE and PBE0+vdW/beyond-RPA adsorption energies.

On the other hand, recalculating the PBE0+vdW ($p_{\textrm{O}_2}$, $T$) phase diagram for Mg$_2$O$_x$ (Fig.~\ref{pt2}) with the experimental O$_2$ binding energy results only in minor changes in the diagram (see Suppl. Material of Ref. \cite{letter} for the modified diagram).

\subsection{High-pressure limit: condensation line}
\label{condens}
In a pure oxygen environment, from Eq. \ref{eqn4}, we see the $\mu_{\textrm{O}}$ variations (i.e., $\Delta\mu_{\textrm{O}}$) is restricted to a finite range and the oxide will decompose into bulk cluster and oxygen, if $\Delta\mu_{\textrm{O}}$ is less than the so called ``O-poor limit".\cite{ms5} Therefore, the oxide is only stable if the following relation is valid:
\begin{equation}\label{Eq:Or1}
F_{\textrm{Mg}_M\textrm{O}_x} < F_{\textrm{Mg}_M}  + x\mu_{\textrm{O}}
\end{equation}
We now rewrite $\mu_{\textrm{O}}$ with respect to $\mu^\textrm{ref}$ as below:
\begin{equation}
 x \mu_{\textrm{O}}=x (\mu_{\textrm{O}}-\mu^\textrm{ref})+x \mu^\textrm{ref}=x (\Delta\mu_{\textrm{O}}+ \mu^\textrm{ref})
\end{equation}
where $\mu^\textrm{ref}$ is equal to $\frac{1}{2}(E^\textrm{DFT}_{\textrm{O}_2}+E^\textrm{ZPE}_{\textrm{O}_2})$, $E^\textrm{ZPE}_{\textrm{O}_2} = h \nu_\textrm{OO}/2$ being the zero point energy of oxygen molecule. Thus, Eq. \ref{Eq:Or1} becomes:
\begin{equation}
x \Delta\mu_{\textrm{O}} > F_{\textrm{Mg}_M\textrm{O}_x}-F_{\textrm{Mg}_M}-\frac{x}{2} \left( E^\textrm{DFT}_{\textrm{O}_2}+E^\textrm{ZPE}_{\textrm{O}_2} \right)
\end{equation}
In the $(T,p_{\textrm{O}_2})$ phase diagrams, the ``O-poor'' limit is the boundary between Mg$_M$ and the adjacent Mg$_M$O$_x$ region(s), with $x>0$. For the shown case of Mg$_2$O$_x$ (Fig. \ref{pt2}) , one can find the border between Mg$_2$ and Mg$_2$O$_1$, where two different isomers are stable in two different regions of the $T,p_{\textrm{O}_2})$ plane, hence the two colors.

The ``O-rich limit"\cite{ms5}, $\Delta\mu_{\textrm{O}}=0$, refers to a condition where oxygen gas is in equilibrium with O$_2$ droplets condensed on the clusters. This leads to the other restriction:
\begin{equation}
  \Delta\mu_{\textrm{O}}<0
\end{equation}
Combining the above two equations we get the range of $\Delta\mu_{\textrm{O}}$:
\begin{equation}
F_{\textrm{Mg}_M\textrm{O}_x}-F_{\textrm{Mg}_M}-\frac{x}{2} \left(E^\textrm{DFT}_{\textrm{O}_2}+E^\textrm{ZPE}_{\textrm{O}_2}\right) < x \Delta \mu_{\textrm{O}} < 0
\end{equation}
From Eq. \ref{eqn13}, we get the another expression for $\Delta \mu_{\textrm{O}}$ as a function of $T$ and $p_{\textrm{O}_2}$:
\begin{equation}
\begin{split}
\Delta\mu_{\textrm{O}}(T,p_{\textrm{O}_2}) = \frac{1}{2} \left( \mu_{\textrm{O}_2}(T,p_{\textrm{O}_2}) - E^\textrm{DFT}_{\textrm{O}_2} - \frac{h \nu_\textrm{OO}}{2} \right)
\end{split}
\end{equation}
I.e., by using $\mu_{\textrm{O}_2} (T,p_{\textrm{O}_2})$ as expressed in Eq. \ref{eqn13}:
\begin{equation}
\begin{split}
\Delta\mu_{\textrm{O}}(T,p_{\textrm{O}_2})  =~
& \frac{1}{2}\left [-k_\textrm{B}T \ln \left[ { \left( \frac{2\pi m}{h^2} \right) }^{\frac{3}{2}} (k_\textrm{B}T)^{\frac{5}{2}} \right] \right. +\\
& +k_\textrm{B}T \ln p_{\textrm{O}_2} -k_\textrm{B}T \ln \left( \frac{8 \pi^2 I_A k_\textrm{B}T}{h^2} \right) +\\
& + k_\textrm{B}T \ln \left[ 1- \exp \left( -\frac{h\nu_\textrm{OO}}{k_\textrm{B}T} \right) \right] + \\
& \left. -~k_\textrm{B}T \ln \mathcal{M} +k_\textrm{B}T \ln\sigma \vphantom{{ \left( \frac{2\pi m}{h^2} \right) }^{\frac{3}{2}}} \right]
\end{split}
\label{Eq:deltamu}
\end{equation}
Since, as we see from this above expression, oxygen condensation point depends on the chosen temperature and pressure, the ``O-rich'' line is not a straight  line. The limiting boundary, $\Delta\mu_{\textrm{O}}=0$, is shown as a black line in Fig. \ref{pt2}.

%

\subsection{Coexistence regions and reactivity of the clusters}
In the phase diagram as shown in Fig. \ref{pt2}, the unlabeled regions in sand color are those where clusters with different stoichiometry are found with free-energy difference less than  $3k_\textrm{B}T$.
When the difference in free energy between two structures is $3k_\textrm{B}T$, the ratio of the Boltzmann factors is $\sim 20$, i.e., the system is 20 times more often in the low-energy state that in the high energy one. If these two are the only states, this ratio is reflected in a population of $\sim 5$\% in the high energy state and $\sim 95$\% in the low-energy one. The value of $3k_\textrm{B}T$ was indeed chosen in order to have a significant population of the high energy state.
This is one manifestation of the unusual properties of matter at the (sub-)nanoscale. At the thermodynamic limit, coexistence regions are collapsed to a line, while for systems with few degrees of freedom, these are typically extended regions.
The compositions competing in a given coexistence region are at least those adjacent to such region. However, more compositions can contribute to the population. In order to find them, one has to resort to a more detailed diagram, like those of Figs. \ref{pt2}.

Besides the regions where different stoichiometries compete, we have also marked those region at fixed stoichiometry where two or more isomers compete.

In both cases the dynamic transformation between structures (the transition barriers and rates among structures will be assessed elsewhere, but we have noticed that these barriers are typically quite low for small clusters) may enhance the reactivity of the involved clusters.
If a second reactant (i.e., besides oxygen) is added to the gas phase after equilibrium between Mg clusters and oxygen is achieved, in the coexistence regions the new reactant will find a highly dynamical environment. Extending the observations of Reuter and Scheffler \cite{aiat2} made on extended surfaces, it is therefore arguable that the reactivity of the clusters towards the new reactant will be enhanced in the coexistence regions.

\section{Conclusions}
In summary, we introduce and thoroughly benchmark a methodology for the efficient and accurate scanning of {\em ab initio} free-energy surfaces.

We demonstrate our methodology by studying gas-phase metal-oxide clusters in thermodynamic equilibrium in an oxygen atmosphere.
This choice is representative of a class of systems for which extensive and very accurate knowledge of the potential-energy surface is mandatory for meaningful predictions.

At first, we scan the potential energy surface of the clusters at different stoichiometries by employing a massive-parallel ``cascade'' genetic algorithm (cGA), together with \textit{ab initio} atomistic thermodynamics (\textit{ai}AT). The underlying concept of our cGA is to first perform an exhaustive genetic-algorithm pre-scanning of possible structures at various stoichiometries by means of a computationally inexpensive reactive force field. The low-energy structures found with this scanning are used as initial guess for another genetic-algorithm scanning with fitness function based on density functional theory. Structures are locally minimized at the van-der-Waals corrected PBE level, but the energetics are calculated at higher level, namely with the hybrid functional PBE0+vdW.
This can be done because we have found that, while the energetics differ noticeably between PBE and PBE0, the locally optimized geometries are not very different. A careful validation of the employed functionals is performed by comparing them to the highest level of DFT: renormalized second-order perturbation theory, rPT2 \cite{xinguo2,xinguo3}, here calculated at PBE and PBE0 orbitals.

The cGA framework is thoroughly validated by comparing it with a reliable and unbiased scheme for the sampling of the PES, i.e., replica-exchange molecular dynamics.

We find that only with certain choices of the parameters the cGA scheme is able to always locate the GM.
In particular, we have shown the benefits derived from using mixed schemes for selection, crossover, and mutation operations.
Alternative and possibly more efficient selection, crossover, and mutation operations, in particular beyond the fixed composition constraint are under study in our group.

The cGA output structures are then examined by applying the concept of {\em ab initio} atomistic thermodynamics {\em ai}AT in order to analyze the temperature and pressure dependence of the composition, structure, and stability of the various isomers for each size of the clusters.
In order to exemplify our methodology, we have applied it to the case of small Mg$_M$ clusters in oxygen atmosphere.

The methodology we introduce is a necessary step in order to understand the \mbox{(meta-)stability} of the clusters in view of their employment for heterogeneous catalysis. Here, we show as an example that Mg$_2$O$_x$ cluster display large areas in the $(p,T)$ phase diagrams at various size in which clusters with different compositions and/or structures can coexist. These regions are suggested as promising regions for an enhanced reactivity of the ensemble of clusters.

\newpage

\section*{APPENDIX}

\subsection*{Effect of the functional on Mg$_2$O$_x$ phase diagram}
\label{ssec:Efffunc}
\begin{figure}[th!]
\centering
\vspace{12pt}
\begin{tabular}{@{\extracolsep{\fill}}cc}
\def\subfigcapskip{8pt}
\subfigure[\, ReaxFF]{\includegraphics[width=0.4\textwidth,clip]{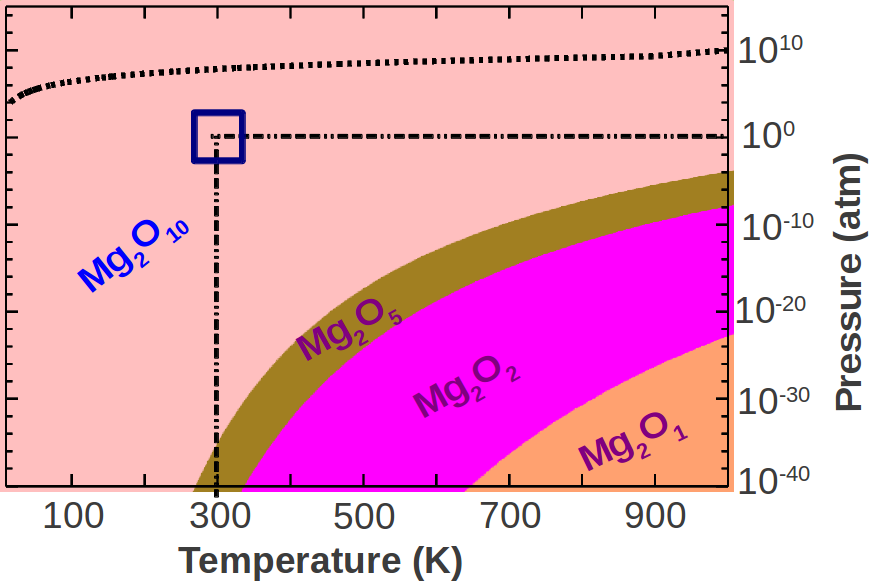}} &
\def\subfigcapskip{8pt}
\subfigure[\, PBE+vdW]{\includegraphics[width=0.4\textwidth,clip]{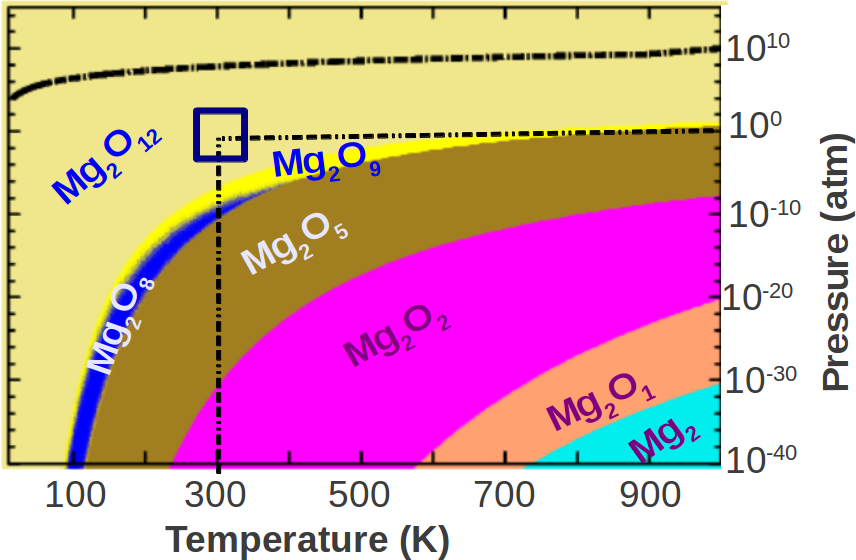}} \\
\def\subfigcapskip{8pt}
\subfigure[\, PBE0+vdW]{\includegraphics[width=0.4\textwidth,clip]{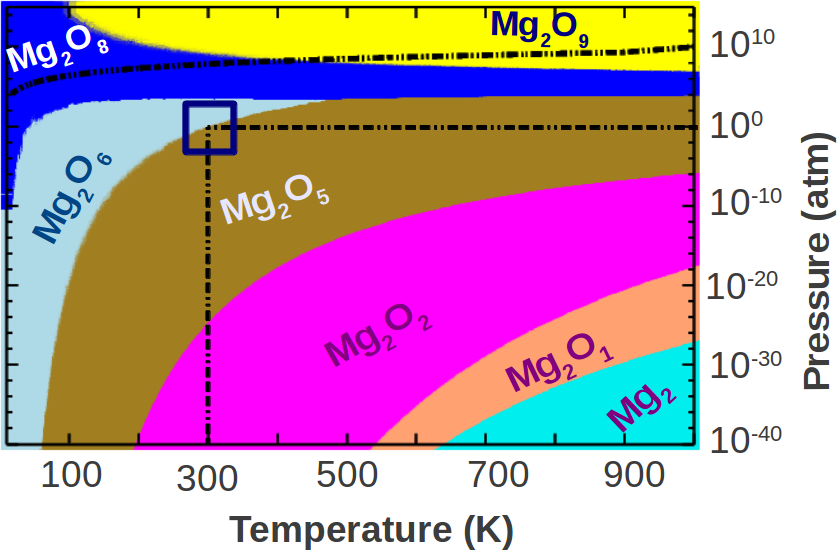}} &
\def\subfigcapskip{8pt}
\subfigure[\, PBE0 (no vdW)]{\includegraphics[width=0.4\textwidth,clip]{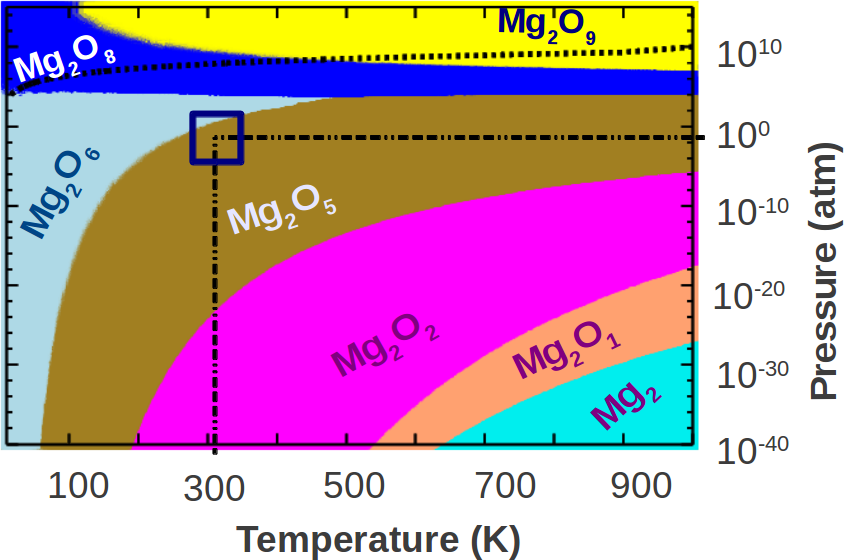}} \\
\def\subfigcapskip{8pt}
\subfigure[\, RPA+rSE@PBE]{\includegraphics[width=0.4\textwidth,clip]{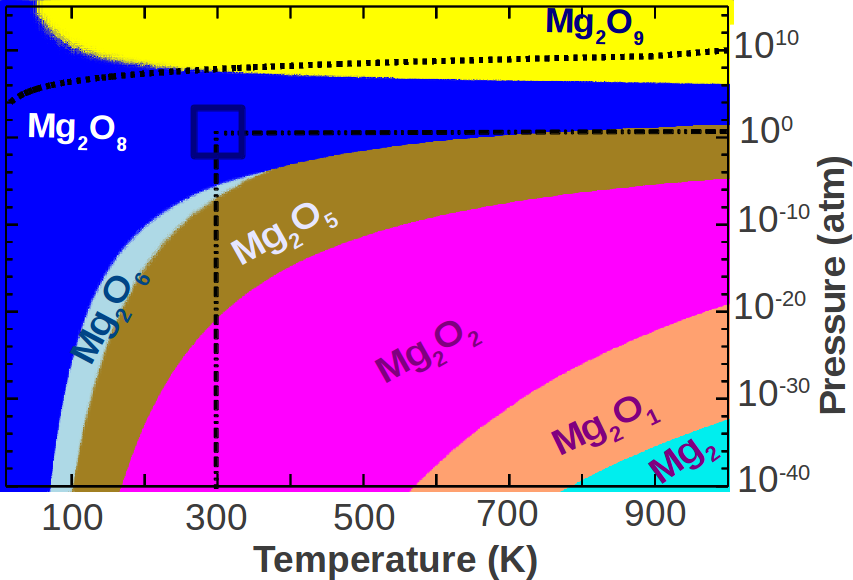}} &
\def\subfigcapskip{8pt}
\subfigure[\, rPT2@PBE]{\includegraphics[width=0.4\textwidth,clip]{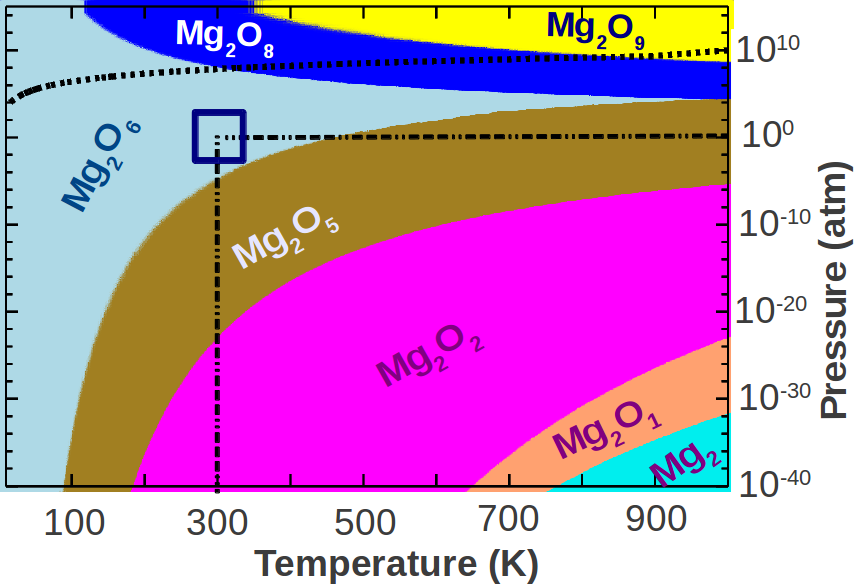}} \\
\end{tabular}
\caption{$(T, p_{\textrm{O}_2})$ phase diagrams of Mg$_2$O$_x$ at (a) ReaxFF, (b) PBE, (c) PBE0+vdW, (d) PBE0 (without vdW), (e) RPA+rSE@PBE and (f) rPT2@PBE levels of theory. For the reaxFF phase diagram, the geometries were optimized with reaxFF and (binding) energies are evaluated with reaxFF. For all the other phase diagrams the geometries are optimized with PBE+vdW, while energies are evaluated with the functional reported in the label of the diagram. The square encompasses the region around normal conditions ($T$ = 300~K, $p_{\rm O_2}$ = 1~atm) and the dashed-dotted lines are guides for the eyes for identifying the point at normal conditions on the diagram.
}
\label{Fig:efffunc1}
\end{figure}

In Fig. \ref{Fig:efffunc1}, we show a comparison of the phase diagrams for Mg$_2$O$_x$ evaluated by using different levels of theory, i.e., reaxFF, PBE+vdW, PBE0+vdW, PBE0 (no vdW), and beyond-RPA. We find that phase diagrams based on PBE0, PBE0+vdW, and beyond-RPA are quantitatively similar. PBE0 and PBE0+vdW yield almost identical phase diagrams, thus at the considered cluster sizes the vdW is not crucial. However, having in mind to study larger clusters, where vdW corrections are expected to play a bigger role, we decided to adopt the PBE0+vdW functional for all sizes.
The boundaries of the stability regions differ by a maximum of 100~K in temperature and 0.1~eV in chemical potential, but exactly the same compositions and structures are present. In particular, the predicted stable composition and structure around normal conditions ($T$ = 300~K and $p_{\textrm{O}_2} = 1$~atm) is the same. In contrast, phase diagrams based on PBE+vdW and FF are very different from those based on PBE0+vdW and beyond-RPA. PBE+vdW (reaxFF) predicts coexistence, within $3k_\textrm{B}T$,  of stoichiometries  from Mg$_2$O$_5$ (Mg$_2$O$_2$) to Mg$_2$O$_{12}$ (Mg$_2$O$_{10}$) around normal conditions, in sharp contrast to PBE0+vdW and beyond-RPA.
This confirms that the PBE+vdW functional is not sufficient even for a qualitative description of the phase diagram of these systems, whereas PBE0+vdW gives qualitatively and quantitatively performances similar to beyond-RPA.

\subsection*{Effect of translational, rotational, vibrational contributions on Mg$_2$O$_x$ phase diagram}
\begin{figure}[t!]
\centering
\begin{tabular}{@{\extracolsep{\fill}}cc}
\def\subfigcapskip{12pt}
\subfigure[\, Including $F^\textrm{translational}, F^\textrm{rotational}, F^\textrm{vibrational}$]{\includegraphics[width=0.4\textwidth,clip]{new-PT-Mg2Ox-PBE0.png}} &
\def\subfigcapskip{12pt}
\subfigure[\, Including only $F^\textrm{vibrational}$]{\includegraphics[width=0.4\textwidth,clip]{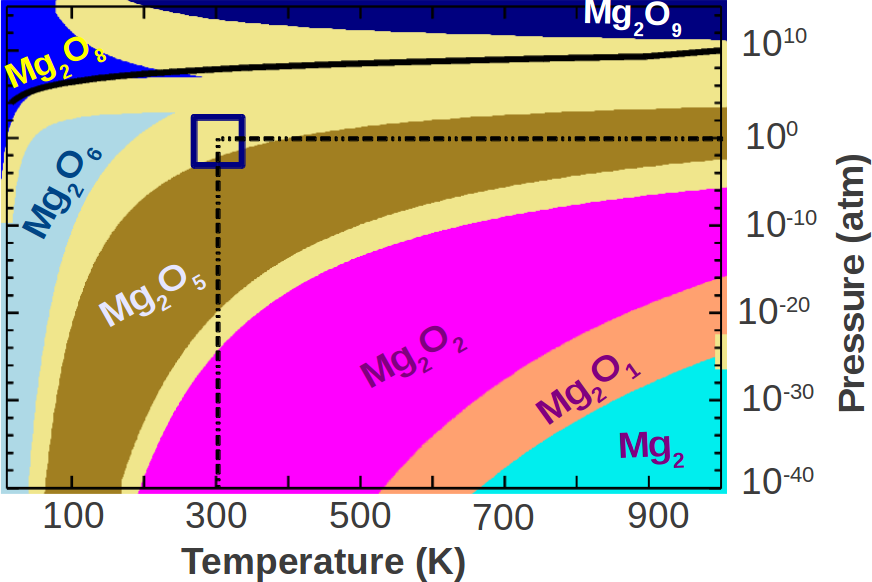}} \\
\def\subfigcapskip{12pt}
\subfigure[\, Including $F^\textrm{translational}, F^\textrm{rotational}$]{\includegraphics[width=0.4\textwidth,clip]{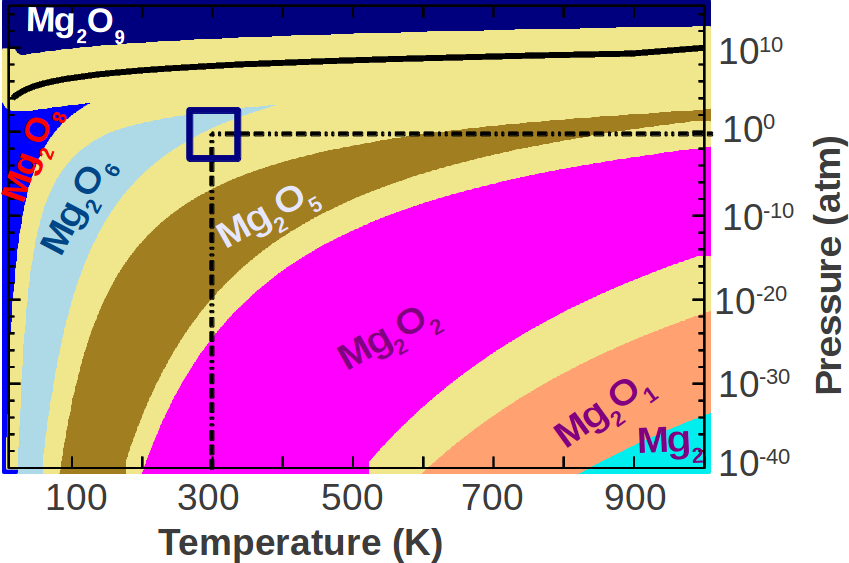}} &
\def\subfigcapskip{12pt}
\subfigure[\, Without $F^\textrm{translational}, F^\textrm{rotational}, F^\textrm{vibrational}$ ]{\includegraphics[width=0.4\textwidth,clip]{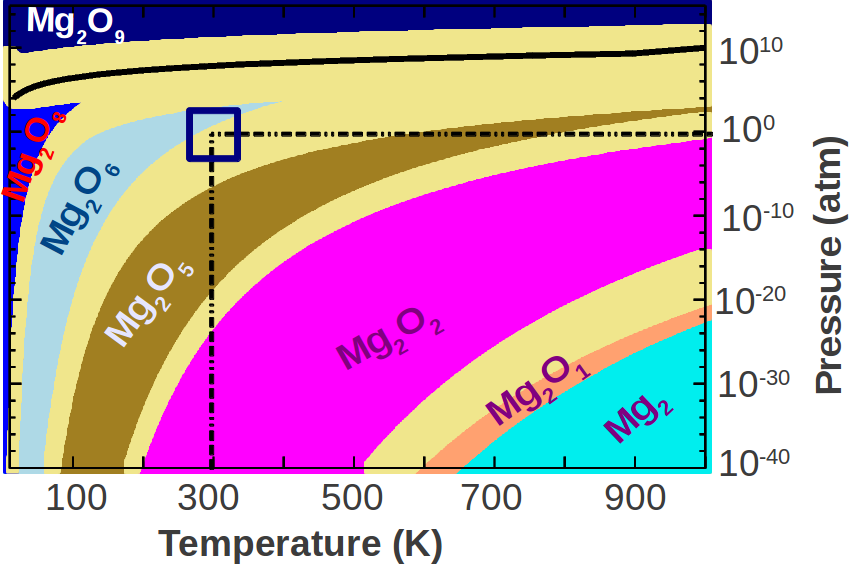}} \\
\end{tabular}
\caption{$(T, p_{\textrm{O}_2}) $ phase diagrams of Mg$_2$O$_x$ with different contributions in the free energy of the clusters (see Eqn. \ref{eqn16}.
The geometries are optimized with PBE+vdW and the electronic energy are calculated using PBE0+vdW. The sand-colored unlabeled regions are regions where different compositions (at least the adjacent ones) coexist (free energy of the coexisting species within 3 $k_\textrm{B}T$, see text).}
\label{Fig:rotvib}
\end{figure}

In Fig. \ref{Fig:rotvib} we show, for the case of Mg$_2$O$_x$ the effect of the inclusion of the  terms $F^\textrm{translational}, F^\textrm{rotational}$, and $F^\textrm{vibrational}$ in the free energy of the clusters (see Eqn.\ref{eqn16}).
We find that neglecting all terms, or including only some of those, results only in slight changes in the phase diagrams, comparable to the differences between PBE0+vdW and beyond-RPA diagrams.

\clearpage

Focusing on the effect of including or not $F^\textrm{vibrational}$, we note that a minor shift of the stability regions of the different compositions.
This shift is such that higher stoichiometries become stable at lower temperatures and higher pressures when $F^\textrm{vibrational}$ is switched on (i.e., stability boundaries shift ``up and to the left'' in the plots). This is due to the fact that the adsorption of one extra oxygen molecule causes a lowering of $F^\textrm{vibrational}$ as new vibrational modes are added to the system.
We also considered the exclusion of $F^\textrm{vibrational}$ for Mg$_{10}$O$_x$ and also in this case the perturbation to the phase diagram was barely visible, but in the same direction as for Mg$_2$O$_x$. Thus, the observation does not change with the size of the cluster.
In this paper, we consider only the case of harmonic contributions to the vibrational free energy. This is not always a good approximation \cite{lmg1}, in particular at higher temperatures. Strictly relying on harmonic free-energy is thus a possible limitation of our methodology: An efficient strategy when large anharmonic contributions are present, but the structures do not melt, is to evaluate the excess anharmonic free energy through thermodynamic integration, as outlined in Ref. \cite{letter}. This approach will be analyzed in detail elsewhere.

\begin{figure}[h!]
\begin{tabular}{@{\extracolsep{\fill}}cc}
\subfigure{\includegraphics[width=0.4\textwidth,clip]{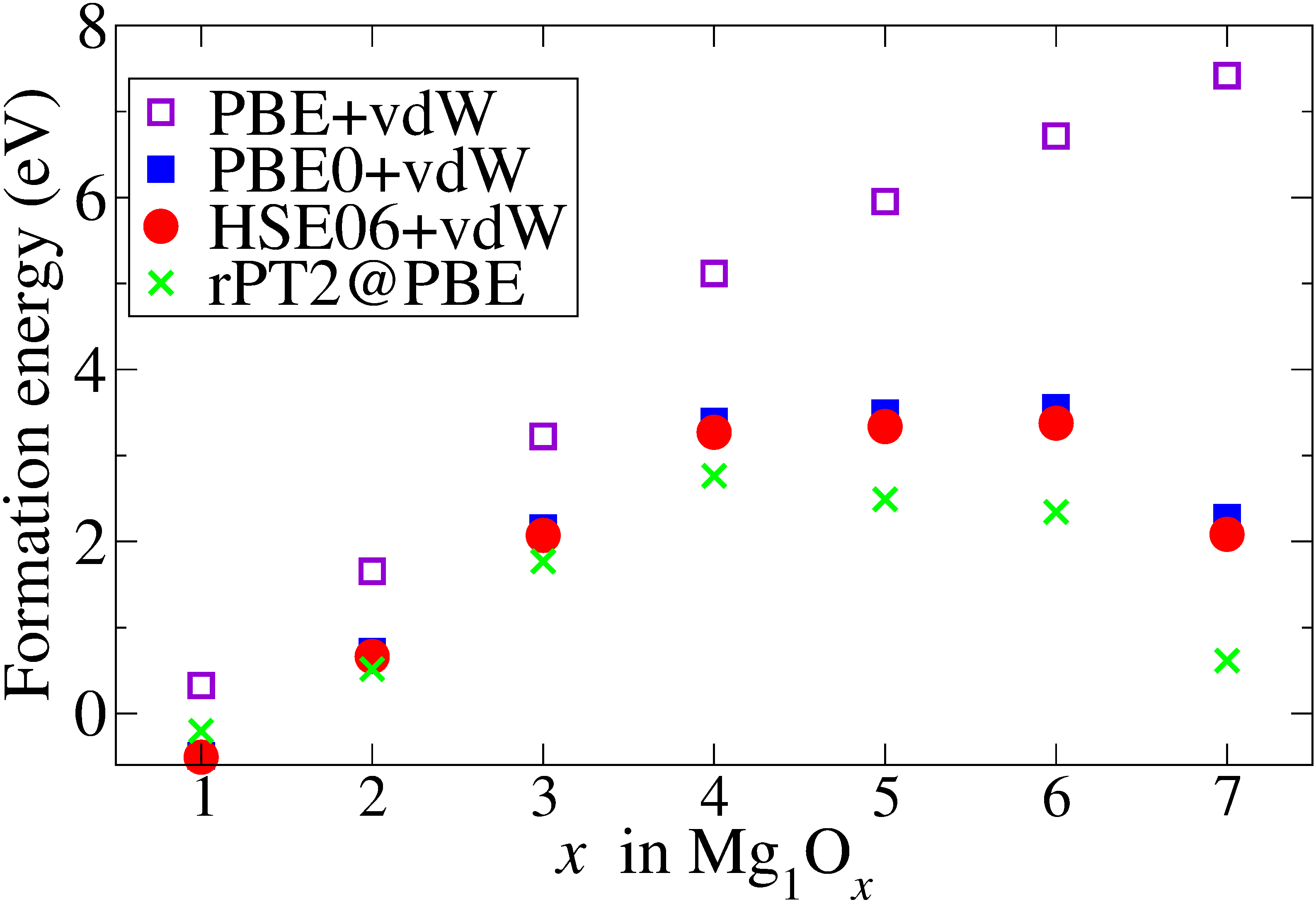}} &
\subfigure{\includegraphics[width=0.4\textwidth,clip]{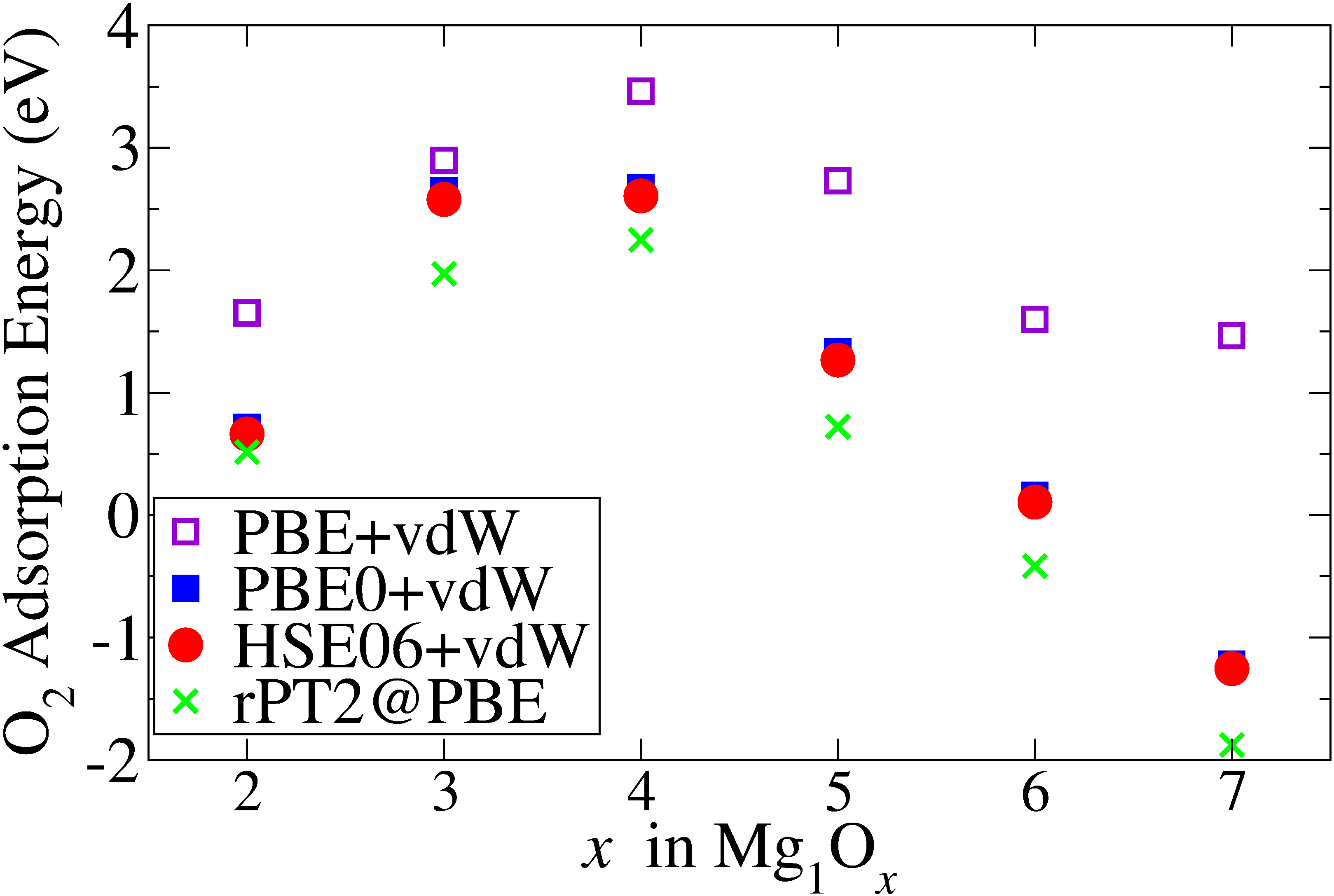}}\\
\end{tabular}
\caption{Formation energy (top) and O$_2$-adsorption energy (bottom) of MgO$_x$ cluster using different xc-functionals, referred to the experimental value of O$_2$ binding energy (5.21 eV [67]). The formation energy of MgO$_x$ is defined similarly to Fig. \ref{be1} (left), but here the binding energy of O$_2$ is taken from the experimental value for all functionals: $ \Delta \tilde{E}^\textrm{form} = E^F_{\textrm{MgO}_x} - E^F_{\textrm{Mg}} - \frac{x}{2} \tilde{E}_{\textrm{O}_2}$, where $\tilde{E}_{\textrm{O}_2} = 2 E^F_{\textrm{O}} + E^{b,\textrm{exp}}_{\textrm{O}_2}$.
$E^F_{\textrm{O}}$ is the total energy of one oxygen atom using the functional $F$ ($F =$ PBE+vdW, PBE0+vdW, HSE06+vdW, rPT2@PBE0), and $E^{b,\textrm{exp}}_{\textrm{O}_2}$ is the experimental binding energy of O$_2$. }
\label{Fig:O2expBE}
\end{figure}

\subsection*{Effect of O$_2$-binding-energy accuracy}
\label{O2bind}

As can be seen in Fig.~\ref{be1}, for lower O$_2$-coverage the difference between PBE+vdW and PBE0+vdW/beyond-RPA energies of O$_2$ adsorption on Mg and  MgO$_2$ is small despite the error in the O$_2$ binding energy (6.23 eV for PBE+vdW, 5.36 for PBE0+vdW, 5.02 for RPA+rSE@PBE, 4.94 for RPA+rSE@PBE0, 4.59 for rPT2@PBE, and 4.42 for rPT2@PBE0, versus 5.21 eV experimental~\cite{exp-o2}). This can be explained by the cancellation of the error for the clusters. Indeed, adsorption of O$_2$ on Mg and MgO$_2$ does not lead to formation or breaking of O-O bonds. While this is also true for MgO, we find that MgO itself has higher atomization energy and ionization potential at PBE level, which leads to a weaker binding to O$_2$.  For clusters with $x\geq 5$, correction of the O$_2$ binding energy error (Fig. \ref{Fig:O2expBE}) increases the difference between PBE and PBE0+vdW/beyond-RPA adsorption energies.



\end{document}